\documentclass{pasa}%

\usepackage{graphicx}

\title[A Versatile High-Precision Polarimeter]{HIPPI-2: A Versatile High-Precision Polarimeter}

\author[Bailey et al.]{Jeremy Bailey$^1$\thanks{j.bailey@unsw.edu.au}, Daniel V. Cotton$^{1,2,3}$, Lucyna Kedziora-Chudczer$^{1,3}$, Ain De Horta$^2$, Darren Maybour$^2$ 
\affil{$^1$School of Physics, UNSW Sydney, New South Wales, 2052, Australia}%
\affil{$^2$Western Sydney University, Locked Bag 1797, Penrith-South DC, NSW 1797, Australia}
\affil{$^3$Centre for Astrophysics, University of Southern Queensland, Toowoomba, Queensland. 4350. Australia}
}%

\jid{PASA}
\doi{10.1017/pas.\the\year.xxx}
\jyear{\the\year}

\usepackage{aas_macros}
\usepackage{hyperref}
\hypersetup{colorlinks,citecolor=blue,linkcolor=blue,urlcolor=blue}

\hypersetup{draft}

\newcommand\0{\phantom{0}}
\newcommand\arcsec{$^{\prime\prime}$}
\newcommand\sdssg{g\,$^{\prime}$}
\newcommand\sdssr{r\,$^{\prime}$}

\begin{document}

\begin{frontmatter}
\maketitle

\begin{abstract}
We describe the High-Precision Polarimetric Instrument-2 (HIPPI-2) a highly versatile stellar polarimeter developed at the University of New South Wales (UNSW). Two copies of HIPPI-2 have been built and used on the 60-cm telescope at Western Sydney University's (WSU) Penrith Observatory, the 8.1-m Gemini North Telescope at Mauna Kea and extensively on the 3.9-m Anglo-Australian Telescope (AAT). The precision of polarimetry, measured from repeat observations of bright stars in the SDSS \sdssg{} band, is better than 3.5 ppm (parts per million) on the 3.9-m AAT and better than 11 ppm on the 60-cm WSU telescope. The precision is better at redder wavelengths and poorer in the blue. On the Gemini North 8-m telescope the performance is limited by a very large and strongly wavelength dependent telescope polarization that reached 1000's of ppm at blue wavelengths and is much larger than we have seen on any other telescope.
\end{abstract}

\end{frontmatter}

\section{INTRODUCTION }
\label{sec:intro}

Polarization measurements of stars using ground-based telescopes can be made with very high levels of precision. As a differential measurement polarimetry is not subject to the same atmospheric effects that limit the precision of photometry. Techniques based on rapid modulation using photoelastic modulator technology \citep{kemp81} have enabled the development of stellar polarimeters capable of parts per million levels of precision \citep{hough06,wiktorowicz08,wiktorowicz15}. High precisions  ($\sim$10~ppm) have also been achieved using a double image polarimeter with a rotating wave-plate modulator \citep{piirola14}. 

The High Precision Polarimetric Instrument \citep[HIPPI, ][]{bailey15} used an alternate approach based on a Ferro-electric Liquid Crystal (FLC) modulator. HIPPI was used on the 3.9-m Anglo-Australian Telescope (AAT), was commissioned in 2014 and demonstrated a precision on bright stars of 4.3~ppm in fractional polarization \citep{bailey15}. HIPPI has been successfully used for a range of science programs including surveys of polarization in bright stars \citep{cotton16}, the first detection of polarization due to rotational distortion in hot stars \citep{cotton17a}, studies of the polarization in active dwarfs \citep{cotton17b,cotton19a}, the interstellar medium \citep{cotton17b,cotton19b} and hot dust \citep{marshall16} and some of the most sensitive searches for polarized reflected light from exoplanets \citep{bott16,bott18}.

HIPPI-2 is a redesigned instrument that incorporates a number of improvements based on our experience with -- and extensive use of -- HIPPI, as well as with the compact and lightweight Mini-HIPPI instrument \citep{bailey17}. HIPPI-2 shares with its predecessors the use of FLC modulators, a polarizing beam-splitter prism and photomultiplier-tubes (PMTs) as detectors. However, HIPPI-2 uses a redesigned optical system, a new, largely 3D printed, construction, and a compact low-power electronics system that replaces $\sim$30~kg of rack-mount electronics in the original HIPPI, with a single compact electronics box weighing 1.3~kg. HIPPI-2 provides improvements in optical throughput and observing efficiency. It is sufficiently compact and lightweight to be easily mounted on small telescopes such as the WSU (Western Sydney University) 60-cm telescope, but powerful enough to provide unique capabilities to very large telescopes.

In this paper we describe the HIPPI-2 instrument and its data reduction and analysis techniques, and evaluate its performance using observations on three telescopes: the 60~cm Ritchey-Chretien telescope at WSU's Penrith Observatory, the 3.9-m Anglo-Australian Telescope at Siding Spring Observatory, New South Wales, and the 8.1-m Gemini North telescope at Mauna Kea, Hawaii.
 
\section{INSTRUMENT DESCRIPTION}

\begin{figure}
\includegraphics[width=7.5cm]{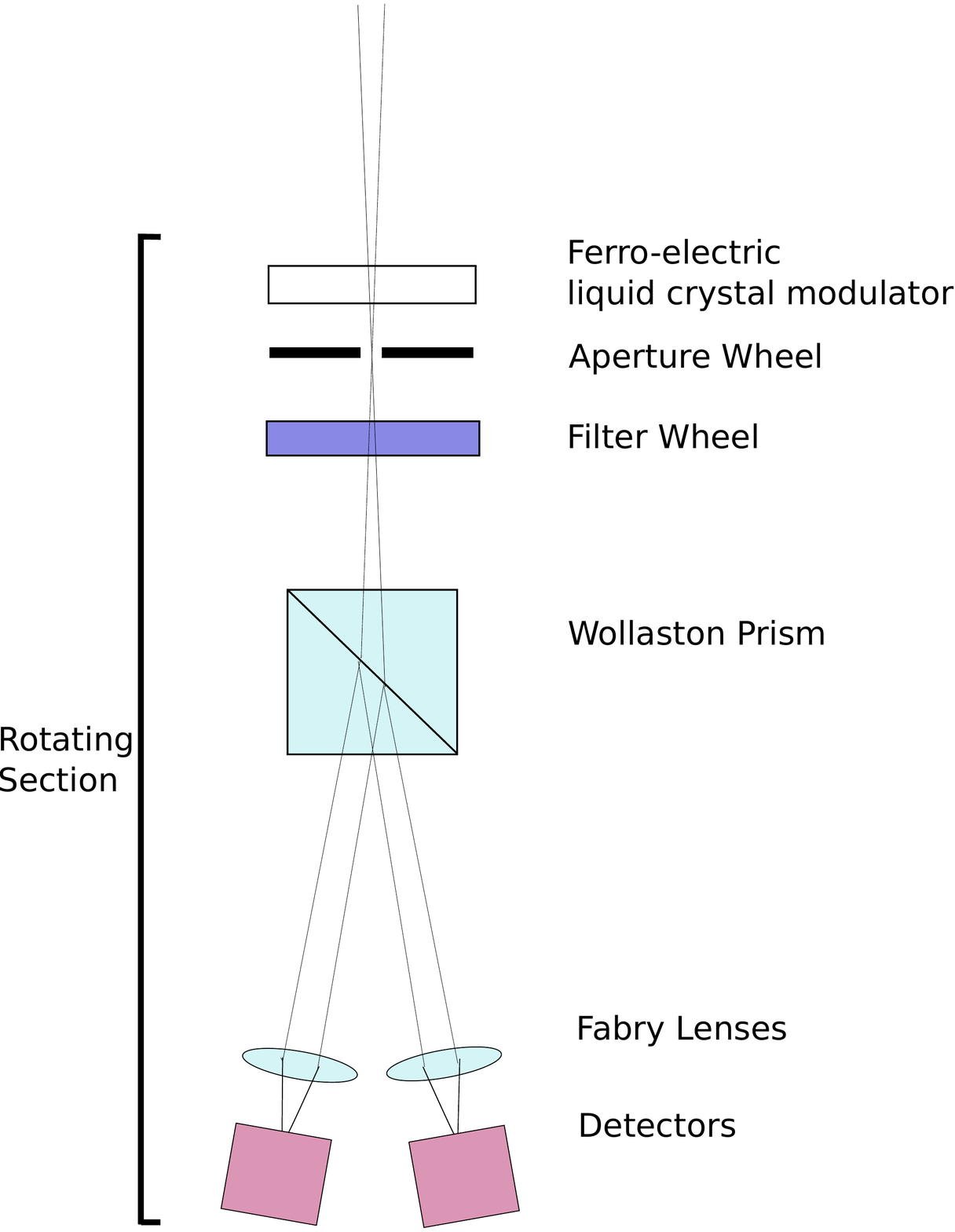}
\caption{Schematic diagram of HIPPI-2 optical system (not to scale)}
 \label{fig1}
\end{figure}

\subsection{Overview}
The optical system of HIPPI-2 is shown in figure \ref{fig1} which is adapted from the similar figure for HIPPI in \citet{bailey15}. HIPPI-2 is designed for a slower input beam (f/16) rather than the f/8 used in HIPPI. This allows the instrument to dispense with the collimating lenses used in HIPPI. The same Fabry lenses (Thorlabs AC127-019A) and Wollaston prism (Thorlabs WP10-A) used in HIPPI are used for HIPPI-2. The rotating section of the instrument is now the whole instrument, rather than just the prism and detectors as in HIPPI. 

The Gemini North telescope has an f/16 focal ratio that matches HIPPI-2. On the AAT it was intended to use HIPPI-2 at the f/15 Cassegrain focus. However due to problems with the coating of the f/15 secondary, it has sometimes been used at the f/8 focus, with a $-$150 mm focal length negative achromatic lens (Edmund Optics 45423) to convert the beam to approximately f/16. On the 60-cm WSU telescope which has an f/10.5 focal ratio, the negative lens is also used giving an effective f/21 beam.

When used, the transmission of the negative lens sets the short wavelength limit of the instrument response, a role otherwise taken by the Fabry lenses, as shown in figure \ref{fig:lenses}\footnote{The transmission data in figures \ref{fig:lenses}, \ref{fig:modulators} and \ref{fig:filters} are available in a public repository at github.com/JbaileyAstro/hippi2. The curve representing the transmission of the Wollaston prism is one measured using a Cary 1E UV-Vis spectrometer for a similarly coated Thorlabs Glan Taylor prism.}.

To minimize telescope polarization due to inclined mirrors HIPPI-2 needs to be mounted at a direct Cassegrain focus. On Gemini North it mounts on the up-looking science port of the Instrument Support Structure to avoid the need to use the science fold mirror. On the AAT HIPPI-2 mounts on the CURE Cassegrain interface unit \citep{horton12}.

\subsection{Ferro-electric liquid crystal modulators}

HIPPI-2 uses FLC modulators operating at 500 Hz for the primary polarization modulation. FLCs are electrically switched half-wave plates. They have a fixed retardation but the orientation of the fast axis can be switched by applying a square wave voltage.

Two different modulators have been used with HIPPI-2. For the 2018 AAT and WSU telescope observations, we used the same MS Series polarization rotator from Boulder Non-linear Systems (BNS), that we previously used with HIPPI. This modulator is driven by a $\pm$5~V square wave signal. The modulation efficiency of the BNS modulator was described by \citet{bailey15}. However, we have found its performance to drift over time requiring re-calibration as described later in section \ref{sec:bandpass}. 

For the Gemini North observations and the 2019 observations with the AAT and WSU telescope we used a 25~mm diameter modulator from Meadowlark Optics (ML) with a design wavelength of 500~nm. This modulator uses a $\pm$9~V square wave drive signal. We can compare the different modulators using the product of their modulation efficiency and transmission over the wavelength range of interest. The modulation efficiency curves are described in section \ref{sec:bandpass}, the transmission of the modulators is shown in figure \ref{fig:modulators}.

The FLCs are temperature sensitive devices and so are mounted in temperature controlled enclosures and operated at a temperature of 25~$\pm$
~0.2~$^{\circ}$C.

\subsection{Filter and Aperture Wheels}

HIPPI-2 includes filter and aperture wheels. HIPPI had a six position filter wheel and only a single fixed aperture. The provision of an aperture wheel with various size apertures, allows the choice of aperture to be optimized for the seeing conditions and background level, and provides a capability to study extended objects such as debris disks and Solar system planets. Table \ref{tab_aperture} lists the standard set of six apertures with the size in arc seconds for Gemini North, the AAT (f/15) and WSU 60-cm (f/21).

\begin{table}
\centering
\caption{HIPPI-2 Apertures}
\begin{tabular}{crrrr}
\hline
Position & Size & Gemini & AAT & WSU \\
         & (mm) & (\arcsec) & (\arcsec) & (\arcsec) \\
\hline
1 & 1.6\0 & 2.6 & 5.7 & 26.2 \\
2 & 2.6\0 & 4.2 & 9.3 & 42.6 \\
3 & 3.6\0 & 5.8 & 12.8 & 58.9 \\
4 & 4.75 & 7.6 & 16.9 & 77.8 \\
5 & 5.65 & 9.1 & 20.1 & 92.5 \\
6 & 7.7\0 & 12.4 & 27.4 & 126.0 \\
\hline
\end{tabular}
\label{tab_aperture}
\end{table}

\begin{table}
\centering
\caption{HIPPI-2 Filters}
\begin{tabular}{llll}
\hline
Position & Name & $\lambda$ (nm) & Notes \\
 \hline
1*& 650LP & >650 & Long Pass Filter \\
  & U & 337-392  & Omega Optics Bessell   \\
2 & V & 480-590  & Omega Optics Bessell   \\
3 & Clear &  & No Filter  \\
4 & \sdssr & 562-695  & Astrodon Gen 2  \\
5 & 500SP & <500 & Short Pass Filter  \\
6 & 425SP & <425 & Short Pass Filter \\
7 & Blank &      &  \\
8 & \sdssg{}     & 401-550 & Astrodon Gen 2 \\
\hline
\end{tabular}
\begin{flushleft}
Notes:
* Two filters have been used in position 1. \\
\end{flushleft}
\label{tab_filter}
\end{table}

The HIPPI-2 filter wheel has eight positions and can accept 25 or 27~mm diameter circular filters. The set of filters used so far are listed in table \ref{tab_filter}. The Blank position in the filter wheel allows measurements of the dark current of the detectors. The SDSS \sdssg{}  and \sdssr{} filters used in HIPPI-2 are generation 2 filters from Astrodon Photometrics and have substantially higher peak transmission and squarer responses than the filters used in HIPPI. The transmission profile of each filter is shown in figure \ref{fig:filters}. It can be seen that the two shortpass filters cut-off at around 300~nm and also have some transmission at wavelengths greater than 650~nm. We use two different detectors with HIPPI-2 (described in the next section), for the one with the bluer response, the longer wavelength leaks are inconsequential.

\begin{figure}
\includegraphics[width=8.5cm]{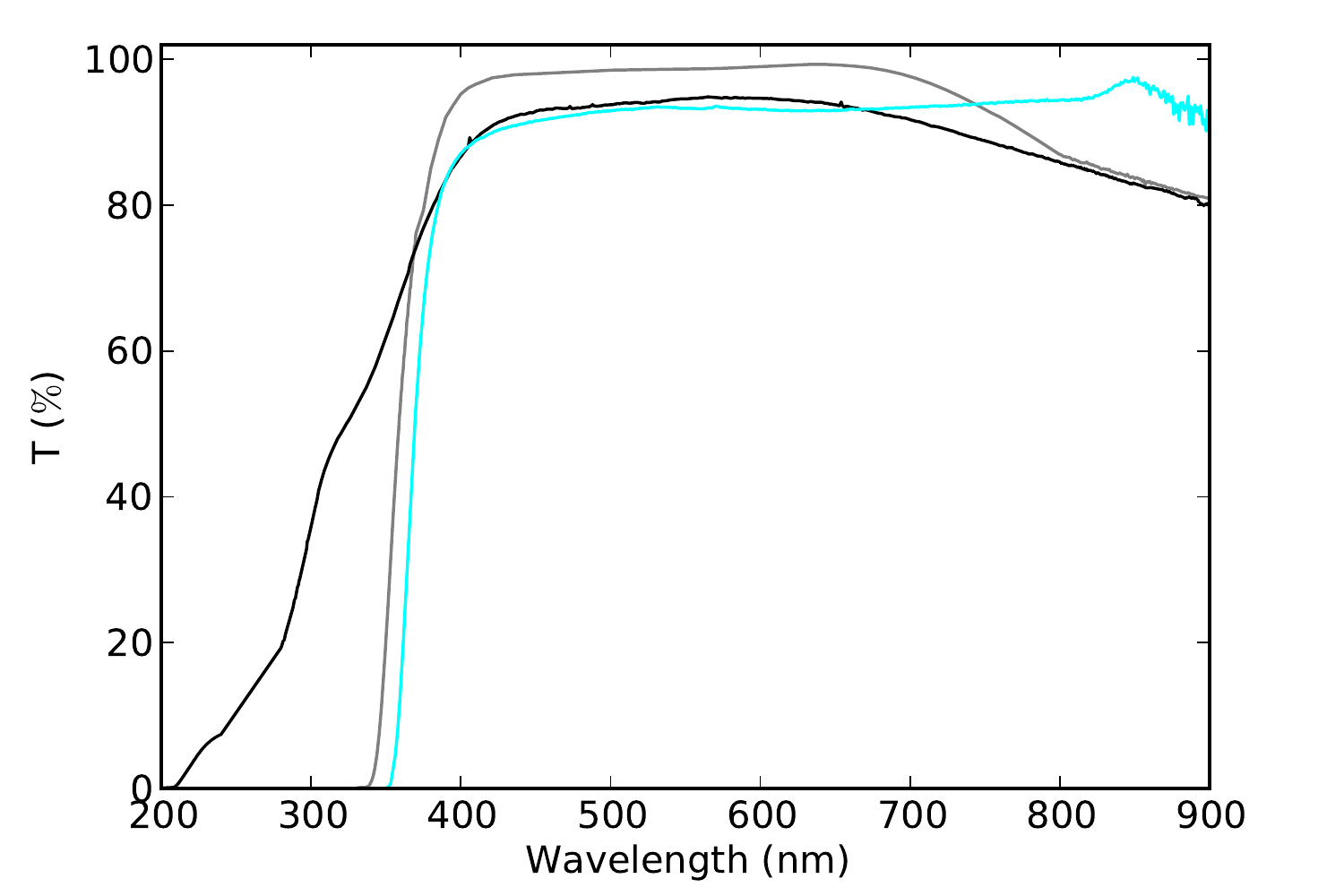}
\caption{The transmission of HIPPI-2 optical components: Wollaston prism (black), Fabry lens (grey), negative achromatic lens (cyan). The transmission data was generated using a combination of manufacturer data and data acquired with a Cary 1E UV-Visible spectrometer.}
 \label{fig:lenses}
\end{figure}

\begin{figure}
\includegraphics[width=8.5cm]{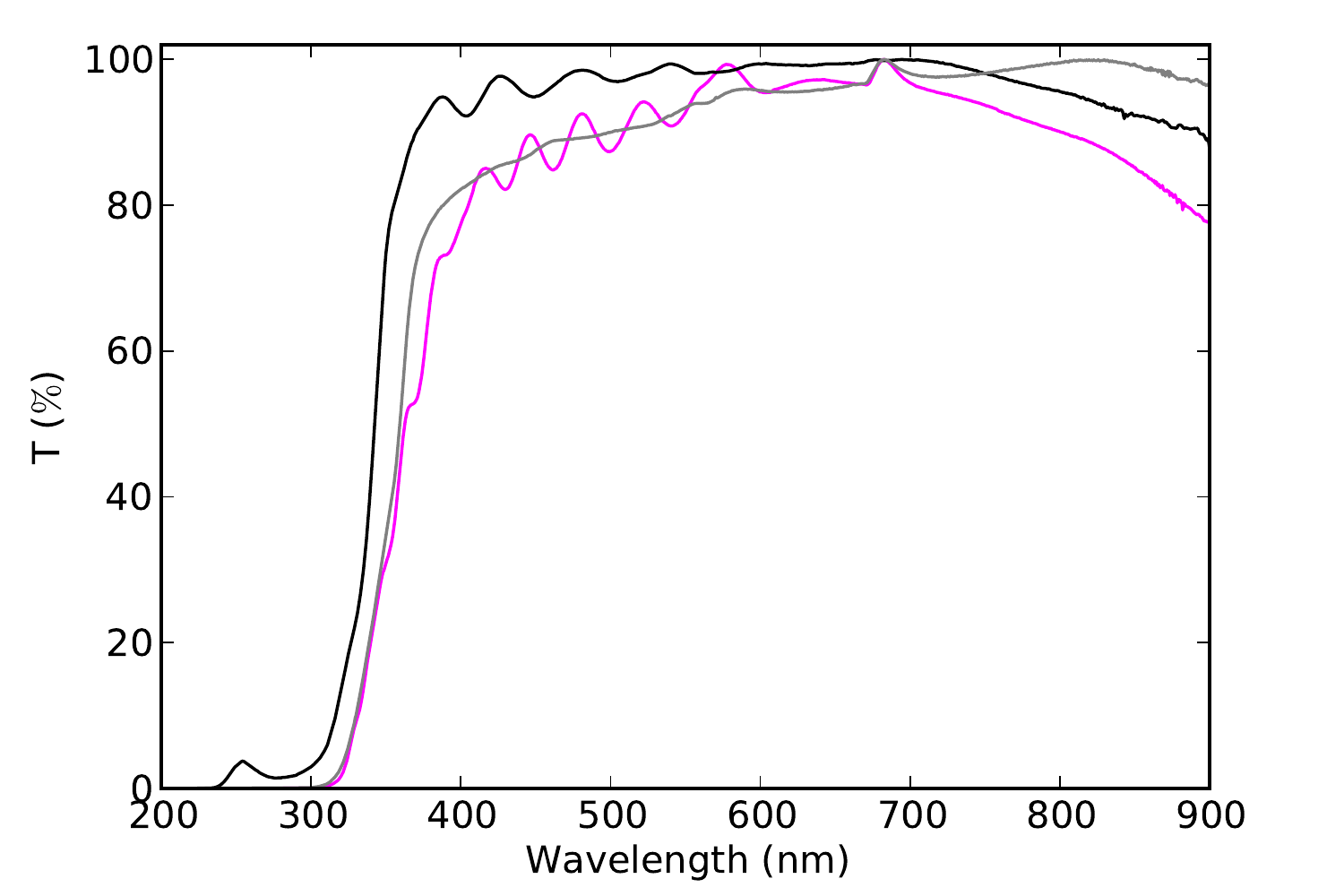}
\caption{The transmission of FLC modulators used with HIPPI-2: ML (black), BNS (grey); and the Micron Technologies (MT, magenta) unit used with HIPPI and Mini-HIPPI. The transmission data was generated using a Cary 1E UV-Visible spectrometer.}
 \label{fig:modulators}
\end{figure}

\begin{figure}
\includegraphics[width=8.5cm]{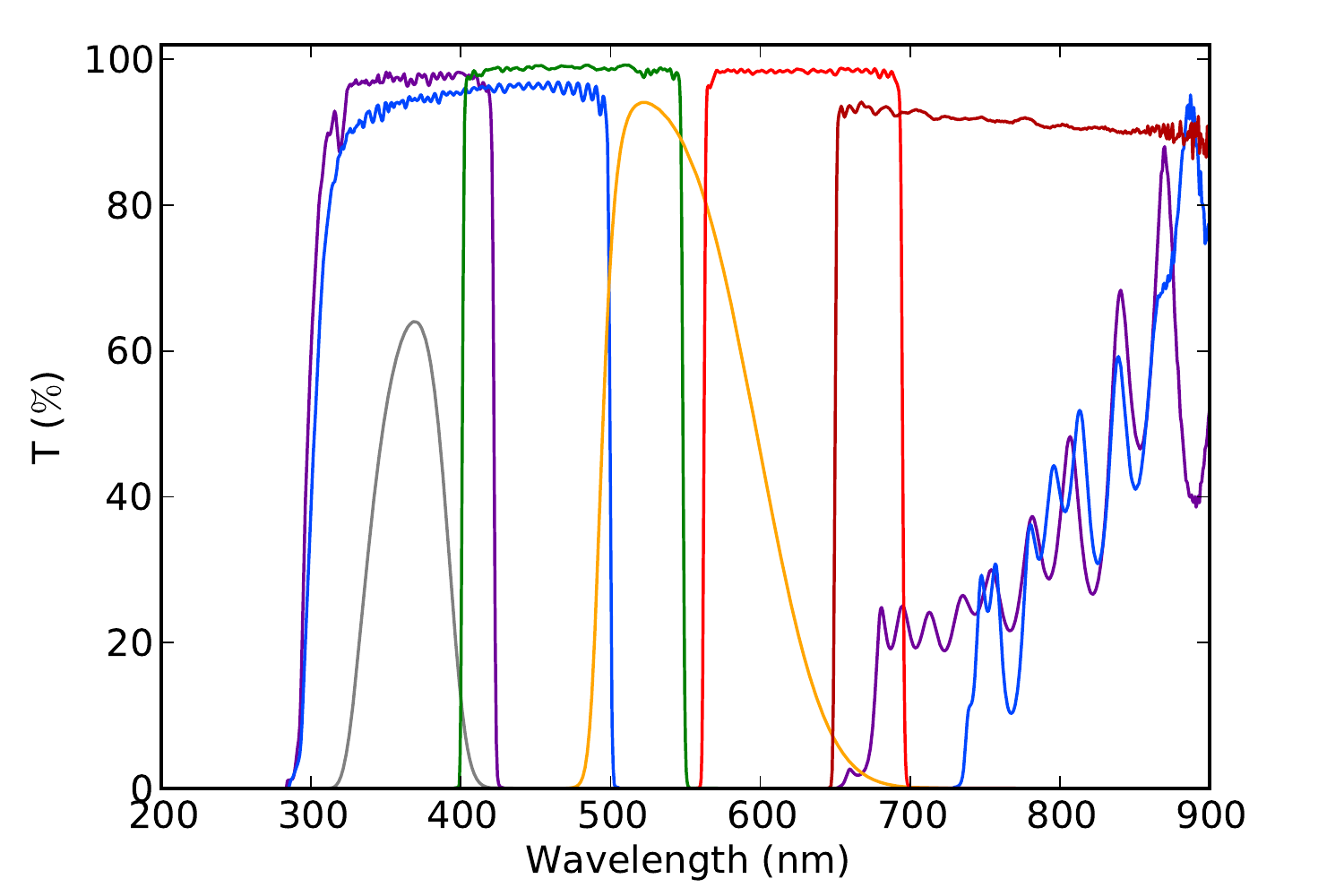}
\caption{The transmission of the HIPPI-2 filters: U (grey), 425SP (violet), 500SP (blue), \sdssg{}  (green), V (orange), \sdssr (red) and 650LP (brown). The U and V band data is manufacturer data, the transmission of the other filters has been determined using a Cary 1E UV-Visible spectrometer.}
 \label{fig:filters}
\end{figure}

\begin{figure}
\includegraphics[width=8.5cm]{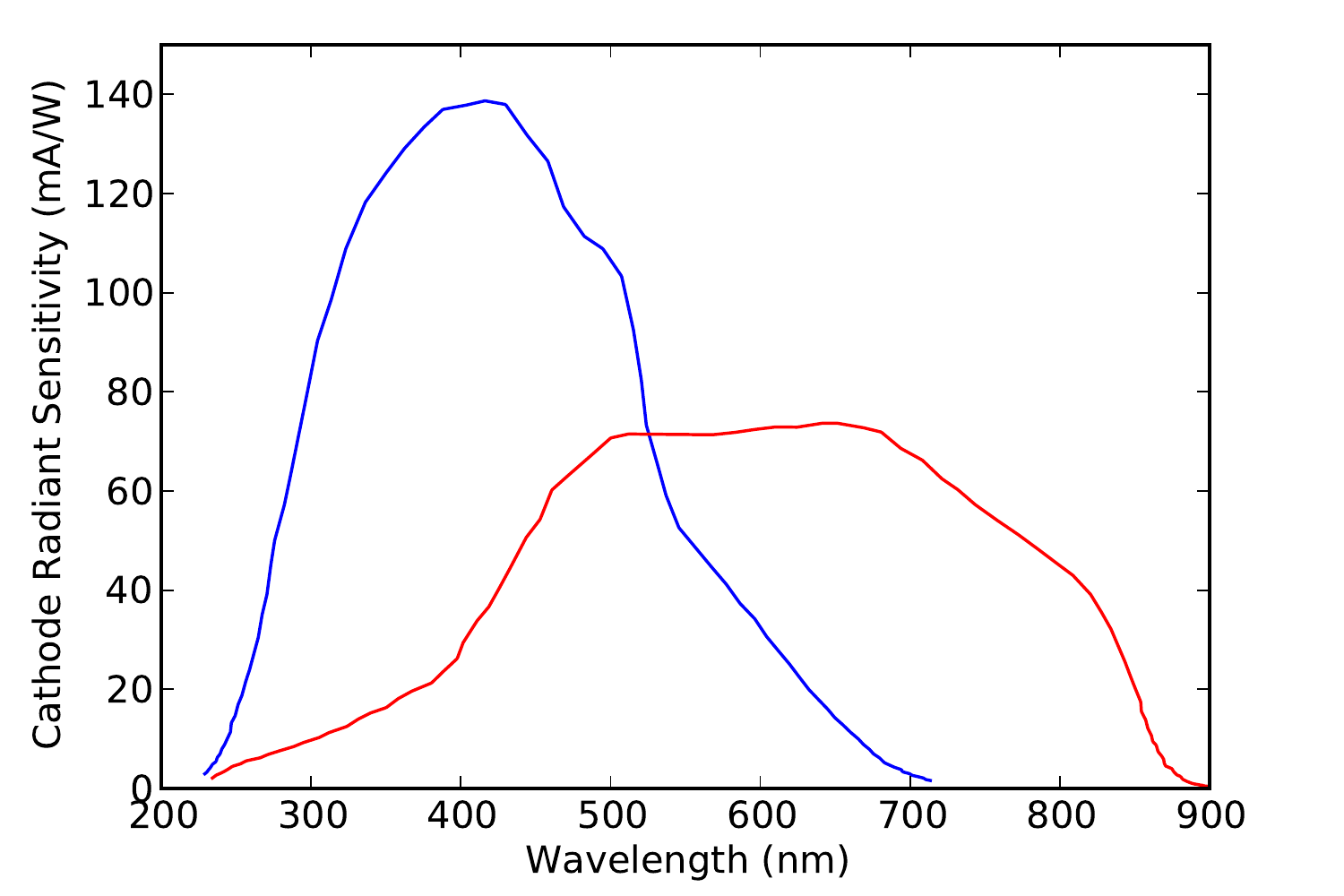}
\caption{The response of the Hamamatsu H10720-210 (blue) and H10720-20 (red) PMTs in mA/W as provided by the manufacturer. Where needed for bandpass calculations the data are interpolated to zero outside of the range of the manufacturer data.}
 \label{fig:detectors}
\end{figure}

\subsection{Detectors}
Following the filter and aperture wheels the Wollaston prism acts as the polarization analyser and splits the light into two beams with a 20$^{\circ}$ separation. A pupil image from each beam is then focused onto the detectors by the achromatic doublet Fabry lenses.

The detectors are compact photomultiplier tube modules (PMTs) containing a metal packaged photomultiplier tube and an integrated high tension (HT) supply. Depending on the application HIPPI-2 can be configured with either blue-sensitive or red-sensitive PMTs. The blue sensitive PMTs (which we denote B) are Hamamatsu H10720-210 modules which have ultrabialkali photocathodes \citep{nakamura10} providing a quantum efficiency of 43\% at 400~nm. The red sensitive PMTs (denoted R) are Hamamatsu H10720-20 modules with extended red multialkali photocathodes. These have a peak quantum efficiency of 19\% at 500~nm and response extending to 900~nm. Figure \ref{fig:detectors} shows the detector response. Switching between blue and red configurations takes 5--10~minutes.

The detector modules are fitted with a transimpedance amplifier to measure the detector current as described in \citet{bailey15}. Both the HT (High Tension) supply voltage and transimpedance gain are remotely switchable, and enable a very high dynamic range. On the AAT HIPPI-2 (like HIPPI) can observe even the brightest stars in the sky while providing close to photon noise limited performance. This ability has proved invaluable in enabling precise calibration and scientific studies of polarization in bright stars \citep[e.g.][]{cotton17a,bailey19}.

\begin{figure}
\includegraphics[width=\columnwidth]{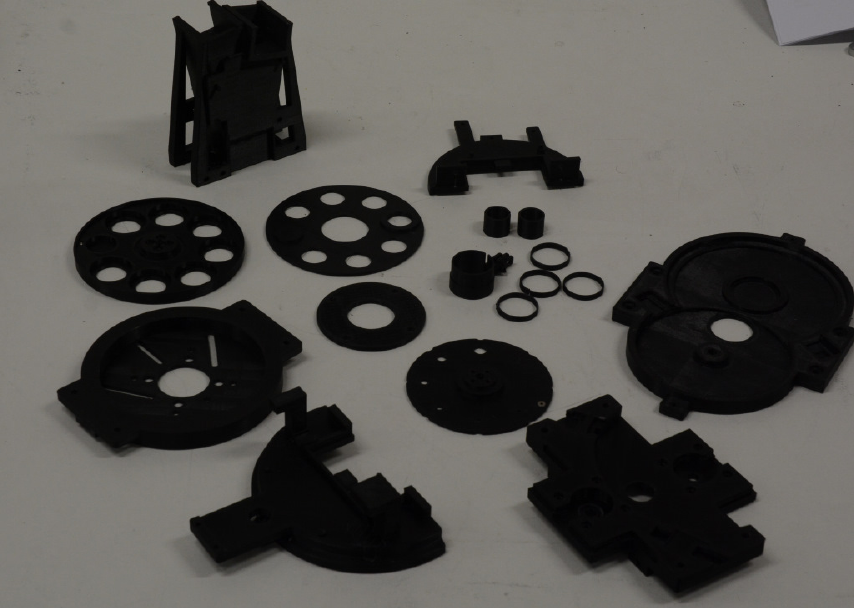}
\caption{3D printed parts for HIPPI-2}
 \label{fig2}
\end{figure}

\subsection{Mechanical Construction}
HIPPI-2 is designed such that the whole instrument can be rotated around the optical axis. The rotation is performed by a Thorlabs NR360S NanoRotator stage. Apart from this rotator and the optical elements already described, the construction of HIPPI-2 is largely by 3D printing. Most of the optical support structure and optical mounts including the filter and aperture wheels were printed in Z-Ultrat material (an enhanced ABS based plastic) on a Zortrax M200 3D printer (see figure \ref{fig2}). Parts for the electronics box were also printed on the same printer. While ABS has a thermal expansion coefficient about 3 times higher than aluminium, the compact design and slow (f/16) optical system mean that the mounting tolerances are not tight and this construction method does not compromise performance.

\begin{figure}
\includegraphics[width=\columnwidth]{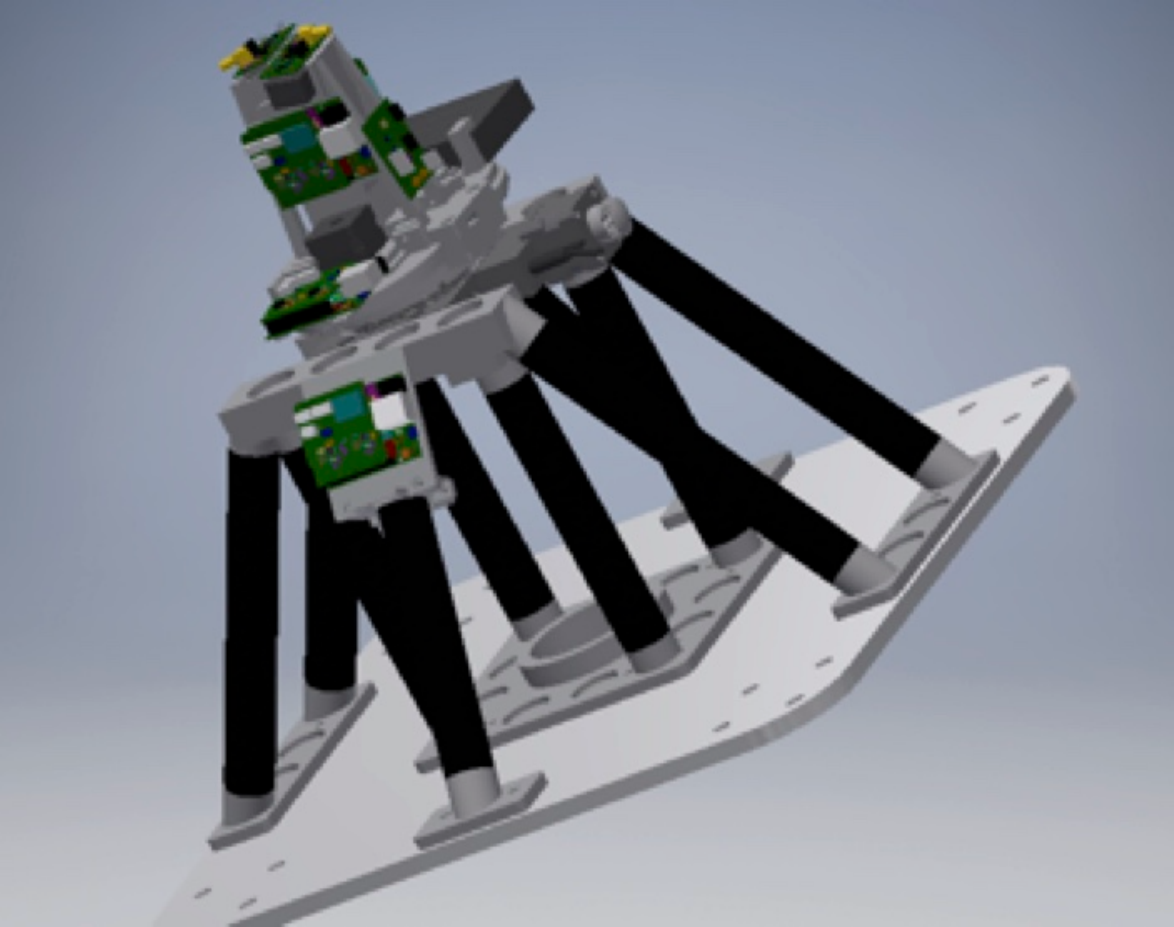}
\caption{HIPPI-2 on its Gemini North Mounting Frame (CAD drawing). Baffling around the optical path is not shown.}
 \label{fig3}
\end{figure}

HIPPI-2 requires a customized mounting for each telescope it is used on. On Gemini North it has to be supported with its aperture 30~cm below the mounting flange at the telescope science port. This is achieved using an aluminium mounting plate, and a support framework of carbon fibre tubing as shown in figure \ref{fig3}, which provides a very strong and stiff structure. The carbon fibre tubes and other components are linked by 3D printed interface pieces (printed on commercial printers in nylon or solid ABS-M30) that are bonded by epoxy to the tubes and bolt to the mounting plate and instrument.

On the AAT the interface to the CURE mounting flange is made from ABS-M30 plastic and manufactured on a Stratasys industrial grade 3D printer. The mounting for the WSU 60-cm telescope uses a mix of 3D printed components and carbon-fibre tubing. 

The only component of HIPPI-2 that required manufacture in a conventional workshop was the aluminium mounting plate for Gemini North. The extensive use of 3D printing, whether on our own printer, or using commercial 3D printing services provides a very fast turnaround, that makes possible a rapid prototyping approach to project development. This helps to reduce costs and speeds up development.

\subsection{Control Electronics and Software}
The control architecture used for HIPPI-2 is based on hardware and techniques developed for the so-called Internet of Things (IoT). Each mechanism or subsystem to be controlled has its own microcontroller which runs a web server and has its own web site that can be used to control and interact with the system. In HIPPI-2, for security reasons, the network is a private network rather than the public Internet.

The microcontroller systems used in HIPPI-2 are EtherTen boards from Australian company Freetronics which use an ATmega 328P CPU and include an Ethernet interface. They are programmed in C++ using the Arduino programming interface. We also experimented with a wireless networked system based on ESP8266 processor boards. While the wireless approach worked well, the radio-quiet requirements of the Mauna Kea site led us to adopt the Ethernet based system.

HIPPI-2 has four subsystems that each have their own microcontroller and web interface. These are the filter and aperture wheels and the FLC temperature controller (these three are all on the rotating part of the instrument) and the instrument rotator (on the fixed part of the instrument). The microcontroller boards and interface electronics are very compact and are mounted on the instrument close to the systems being controlled.

\begin{figure}
\includegraphics[width=\columnwidth]{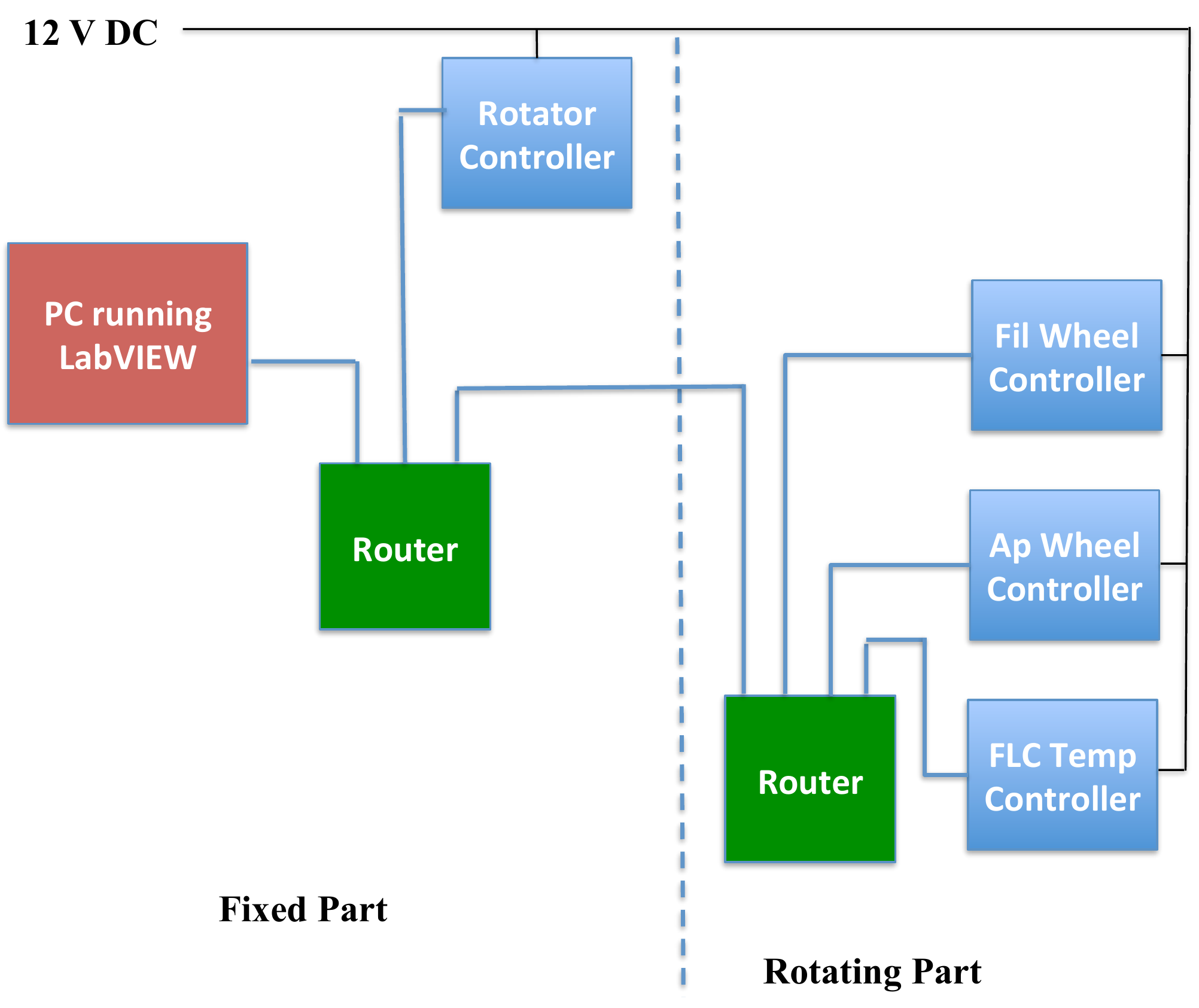}
\caption{HIPPI-2 control architecture showing the Ethernet links between systems. Only a single power cable and one Ethernet cable run between the fixed and rotating parts of the instrument.}
 \label{fig4}
\end{figure}

Ethernet routers (Ubiquiti ER-X with five ports) are mounted on both the fixed and rotating parts of the instrument, and allow the architecture shown in figure \ref{fig4} with only a single Ethernet cable running between the fixed and rotating parts of the instrument. We use a special highly flexible cable (Cicoil DC-500-CA003) for this purpose.

A single 12V DC power supply provides power to all the microcontroller systems as well as the two routers. On board DC-DC converters generate the 5V needed for the microcontroller and any other required voltages. Three of the microcontroller systems (the rotator and filter and aperture wheel controllers) use essentially identical hardware based on a stepper motor driver. The FLC temperature controller implements a proportional-integral (PI) servo to control the drive voltage to a heater, based on feedback from a thermistor temperature sensor.

The microcontroller systems are quite simple devices with relatively slow 8-bit CPUs and lacking an operating system or file system. However, they are small and cheap enough, that we can use one CPU for each mechanism. We do not therefore require them to run complex multi-tasking or real-time software such as is often used at major observatories where a single CPU controls all the functions in an instrument. The software on each microcontroller is about 300--400 lines of fairly straightforward code. Much of the code can be reused between the different controllers. The only user interface required is a web browser. The hardware and software costs of this approach are very low, and these systems have proved very reliable in operation.

\subsection{Data Acquisition}
The data acquisition system for HIPPI-2 is essentially the same as that used for Mini-HIPPI and described by \citet{bailey17}. Two National Instruments USB-6211 data acquisition modules are used to read the data from the detectors as well as providing the drive signal for the FLC modulator and controlling the PMT gain and HT voltage. These modules interface via USB to an Intel NUC miniature PC running Windows 10. This computer also provides the interface to the microcontroller systems as shown in figure \ref{fig4}. The instrument software is adapted from that used for HIPPI and Mini-HIPPI and is written in the LabVIEW graphical programming environment.

The detector signals are read at a 10~$\mu$s sample time, synchronized with the FLC modulation. The data are folded over the modulation period (2~ms for the standard 500~Hz modulation frequency) and written to output files after an integration time of 1--2 seconds.

\subsection{Summary}

HIPPI-2 is a compact and low-cost instrument. On its AAT or WSU mounts the total weight of the instrument is about 4 kg. A compact electronics box weighing 1.3~kg holds the NI interface modules, the computer and the FLC drive electronics and trigger circuitry. The total power requirement for the instrument is about 30~W. The component cost of a complete HIPPI-2 including one set of detectors and filters is about A\$20,000, similar to that of HIPPI, and a little more than that of Mini-HIPPI.

\section{Observing Procedure}
\label{sec:obs}

As with Mini-HIPPI an observation consists of four target measurements at different instrument position angles (PAs): 0, 45, 90 and 135$^{\circ}$. Typically a single sky measurement (with a shorter exposure time) is made at each PA with the same instrument settings. In changeable conditions or on faint targets we sometimes bracket each target measurement between two sky measurements. For bright, or highly polarized targets observed in moonless conditions a single dark measurement can be substituted. Measurements made at the redundant angles (90 and 135$^{\circ}$) are combined with the 0 and 45$^{\circ}$ measurements respectively to enable cancellation of instrumental effects. Instrumental polarization varies with the target's magnitude, the detector voltage settings, and target alignment, so this is an important procedure for obtaining best precision.

HIPPI used the AAT's Cassegrain rotator to rotate the instrument to the four different PAs. With HIPPI-2, the instrument's built-in rotator is used, significantly speeding up observing. HIPPI-2 has two stages of modulation: the electrically driven FLC modulator and the instrument rotation, whereas HIPPI had three stages of modulation, with the third being an instrument back-end rotation swapping the detectors between A and B positions 90$^{\circ}$ apart. We determined from analysis of HIPPI data that this third stage of modulation provided no significant benefit, allowing the simpler system used in HIPPI-2. Eliminating the back-end rotation and using the instrument rather than the telescope rotator --- which reduces the number of target acquisitions from 4 to 1 --- saves on average 5 minutes per observation on the AAT.

\section{Data Reduction and Calibration}
\label{sec:red&cal}

For data taken on an equatorially mounted telescope the data processing is a two-step process, involving first a raw data reduction and then correction. Originally, with HIPPI, the first step was performed by a code written in FORTRAN 77, and the second step using a Microsoft EXCEL spreadsheet. Now both steps are performed using programs written in PYTHON 2.7.5 using elements from the associated packages NUMPY \citep{numpy}, SCIPY \citep{scipy}, MATPLOTLIB \citep{matplotlib}, ASTROPY \citep{astropy1, astropy2}, ASTROPLAN \citep{astroplan} and ASTROQUERY \citep{astroquery}. Although the mathematics of the process is essentially unchanged, rewriting the software has facilitated some improvements of process and enabled better integration between the steps.

\subsection{Raw Data Reduction}
The raw data reduction has three stages: dark and/or sky subtraction; the application of a Mueller matrix model to determine $I$, $Q$ and $U$ for each measurement; and combining the measurements for PA 0, 45, 90 and 135 degrees to produce the raw observation. 

\subsubsection{Dark/Sky Subtraction}
At each PA a sky measurement typically consists of 40 1~s integrations. At 500 Hz operation each integration is made up of 200 modulation points. For each modulation point an average is calculated from all the integrations, and the resulting average integration subtracted from the target measurement point by point. A dark subtraction uses the same procedure. Lab-based dark measurements made for each detector HT voltage and gain setting are subtracted from each target and sky measurement by default; this isn't really necessary when a sky subtraction is applied, but is useful as a monitor of the sky conditions.

\subsubsection{Mueller Matrix Model}
Mathematically the reduction procedure for HIPPI-2 is identical to that of HIPPI as described by \citet{bailey15}. We can describe the instrument by a 4 by 4 Mueller matrix $\mathbf{M}$ that relates the output Stokes vector
$\mathbf{s_{out}}$ to  the input Stokes vector $\mathbf{s_{in}}$ through: 

\begin{equation}
\label{eqn_mull}
\mathbf{s_{out}} = \mathbf{M} \mathbf{s_{in}}
\end{equation}

The Mueller matrix for the instrument is simply the product of the Mueller matrices for its optical components as described by \citet{bailey15}. The Mueller matrix $\mathbf{M}$ is not a constant but varies through the modulation cycle as the modulator properties change.

We can also define a system matrix $\mathbf{W}$. The system matrix is an N by 4 matrix, where each row is a state of the system, corresponding to a single data point in the modulation curve. Multiplying the input Stokes vector by the system matrix gives the vector $\mathbf{x}$ of $N$ observed intensities seen at the detector during the modulation cycle (where $N$ = 200 is the number of data points in our 500 Hz modulation cycle). It can be seen that each of the N rows of $\mathbf{W}$ is the top row of the  Mueller matrix corresponding to that state of the instrument:

\begin{equation}
\label{eqn_sys}
\mathbf{x} = \mathbf{W} \mathbf{s_{in}}
\end{equation}

The system matrix depends on how the waveplate angle and depolarization of the modulator vary through the modulation cycle and we determine this through a laboratory calibration procedure in which we feed polarized light of known polarization states (using a lamp and polarizer) into the instrument for a full rotation of the polarizer in 10 to 20 degree steps.

We can then invert equation \ref{eqn_sys} to give:

\begin{equation}
\label{eqn_red}
\mathbf{s_{in}} = \mathbf{W}^+ \mathbf{x}
\end{equation}

where $\mathbf{W}^+$ is the pseudo-inverse of $\mathbf{W}$, which is calculated numerically. This gives the source Stokes parameters $\mathbf{s_{in}}$ in terms of the observed 
modulation data $\mathbf{x}$.

Further details of the procedure can be found in \citet{bailey15}.

\subsubsection{Combining Measurements to Produce an Observation}
The final Stokes parameters for an observation are determined by combining the measurements for the four instrument PAs. This step has the effect of cancelling out instrumental polarization effects. Only the on-axis Stokes parameter determinations are used, so 0 and 90$^\circ$ contribute to $Q_i/I$, and 45 and 135$^\circ$ to $U_i/I$. The average of all four measurements contribute to a determination of the I Stokes parameter.

\subsection{Correction}
\label{sec:correction}
The correction step involves three processes: the application of a bandpass model, described in section \ref{sec:bandpass}, to scale the polarization magnitude to account for the modulation efficiency of the instrument; a rotation of the co-ordinate frame based on observations of high polarization standards; and subtraction of an offset in $q=Q/I$ and $u=U/I$ associated with the telescope polarization (TP) -- determined by observations of low polarization standards.

\subsubsection{Bandpass Model and Modulator Calibration}
\label{sec:bandpass}

A bandpass model is used to make an efficiency correction and determine the effective wavelength of each observation. The bandpass model used for HIPPI-2 is based on that of HIPPI \citep{bailey15}, and PlanetPol \citep{hough06} before it, but has been rewritten in PYTHON 2 and is extremely versatile. The same code may be used for any combination of source, atmosphere, photosensor, modulator and transmitting (or reflecting) optical components. The bandpass model is integrated into the data processing pipeline, but can also be run independently from the command line or called as a routine in other code enabling full bandpass fitting for science or calibration purposes (e.g. \citealp{cotton19b}).

The effective wavelength is calculated by the bandpass model as \begin{equation}\lambda_{\rm eff}=\frac{\int \lambda S(\lambda)d\lambda}{\int S(\lambda)d\lambda},\end{equation} where $\lambda$ is the wavelength, and $S(\lambda)$ is the relative contribution to the output detector signal as a function of wavelength. In basic terms $S(\lambda)$ includes the product of the photocathode radiant sensitivity (in mA/W) and the source spectral energy distribution (SED) as attenuated by functions describing the atmosphere and optical components of the instrument and telescope. Typically \begin{equation}S(\lambda)=F_{\star} T_{atm} R_{pM} R_{sM} T_{fil} T_{mod} T_{anal} T_{opt} \mathbb{R}_{ph} \end{equation} where every term is a function of $\lambda$ and $F_{\star}$ is the source flux, sometimes modified by reddening, $T_{atm}$ the atmospheric transmission, $R_{pM}$ and $R_{sM}$ the reflectance of the primary and secondary telescope mirrors, $T_{fil}$, $T_{mod}$, $T_{anal}$ and $T_{opt}$ are the transmittance of the filter, modulator, analyser (Wollaston prism) and other optical components in the instrument respectively, and $\mathbb{R}_{ph}$ the radiant sensitivity of the photosensor.

By default a \citet{castelli&kurucz2004} stellar atmosphere model is used for the SED, and sets the resolution of wavelength sampling\footnote{Optionally the wavelength grid can be changed, in which case the SED is interpolated onto the desired grid.}. Included in the bandpass model's standard library are atmosphere models for dwarfs of spectral type O3, B0, A0, F0, G0, K0, M0 and M5\footnote{$F_\lambda$ units ($\textnormal{erg}\,\textnormal{s}^{-1} \textnormal{cm}^{-2} \AA^{-1}$) are assumed, with the option to convert from other units sometimes used by Kurucz ($\textnormal{erg}\,\textnormal{s}^{-1} \textnormal{cm}^{-2} \textnormal{Hz}^{-1}$).}. For intermediate spectral types two bandpass models are calculated and the results linearly interpolated in subtype; the same models are used for other spectral classes. The data reduction software uses a look-up file to determine the spectral type of the target, if absent from the file the object's details are downloaded from SIMBAD by the software for stellar objects (using astroquery, \citealp{astroquery}) or a Solar spectral type assumed for Solar System objects. For distant targets the model SED can be modified to account for interstellar extinction (reddening) using the empirical model of \citet{cardelli89}. However, most of the targets observed with HIPPI-2 are nearby and by default no reddening is applied.

The Earth atmosphere transmission is based on radiative transfer models pre-calculated using VSTAR (Versatile Software for Transfer of Atmospheric Radiation, \citealt{bailey12}). Like the optical component data, the spectrum is spline interpolated onto the wavelength grid set by the source. For the observing sites used in this work, standard built-in models were used. The SSO and MK built-in models were used for the AAT and Gemini North observations respectively. For WSU the built-in mid-latitude summer (MLS) model adjusted for the altitude of the observatory was used. The transmission is calculated for the airmass at the mid-time of each observation.

The PMT sensitivity (in mA/W) is taken from Hamamatsu data sheets as shown in figure \ref{fig:detectors}.  

In the HIPPI bandpass model the instrumental transmission was determined as a whole. Lab based measurements were made with narrowband filters to estimate the attenuation at blue wavelengths. That procedure lacked precision, and for HIPPI-2 we have taken a different approach. The transmission as a function of wavelength has been determined for each optical component separately, with the instrumental transmission being the product of the components in use. The transmittances of the various optical components of the instrument are shown in figures \ref{fig:lenses}, \ref{fig:modulators} and \ref{fig:filters}. The nominal reflectivity of the telescope mirrors is also accounted for in the same way. Where possible we used a Cary 1E UV-Visible spectrometer to make measurements of the filters, modulators and each of the other optical components in the lab, and supplement this with manufacturer data where that proved difficult. This is a better approach than using only manufacturer's data which may not always cover the full wavelength range of HIPPI-2\footnote{We also identified some unadvertised long wavelength light leaks in the shortpass filters.}. Acquiring data for each of the components individually allows for easy and accurate adjustments when components are swapped. The flexibility of this approach has allowed us to use the one bandpass model for all HIPPI-2 observations with and without the negative achromatic lens, as well as all of our older measurements with HIPPI and Mini-HIPPI.

The modulators are designed to be half-wave retarders at one wavelength only. At other wavelengths the modulation efficiency ($e(\lambda)$) will fall off. The raw polarization measurements therefore need correcting by dividing by the effective efficiency given by
 \begin{equation}e_{\rm eff}=\frac{\int e(\lambda) S(\lambda)d\lambda}{\int S(\lambda)d\lambda}.\end{equation}

For any given modulator $e(\lambda)$ may be determined by either a lab based calibration or through on-sky observations of objects with known polarizations. The bandpass model has a built-in option for modelling interstellar polarization by applying either a Serkowski Law \citep{serkowski75} or Serkowski-Wilking Law \citep{wilking82} to the source\footnote{The bandpass model also allows for the addition of source intrinsic polarization through an input file. Additionally, the intrinisc or interstellar polarization can be rotated arbitrarily to return predictions of $q$ and $u$.}. When the source spectrum and polarization as well as the other contributors to the bandpass are well characterised, we can use a forward model to calibrate the modulator performance with a fitting routine, since \begin{equation} p=\frac{\int p_{is}(\lambda) e(\lambda) S(\lambda)d\lambda}{\int S(\lambda)d\lambda},\end{equation} where $p_{is}(\lambda)$ describes the interstellar polarization of the source.

Prior to its first use, the ML modulator was calibrated in our laboratory using as a source the light from an incandescent bulb -- which we approximate as a blackbody -- collimated and directed through a polarizer to produce 100\% polarized light. Measurements were then made with the installed broadband filters and a number of narrowband (NB) filters. The modulation efficiency is different for high and low polarizations (see appendix \ref{sec:apa}). For a 100\% polarized source, the modulation efficiency is given by 

\begin{equation}\label{eq:e_hp} e(\lambda)=  \frac{{}e_{max}}{2}\left (1+\frac{1-\cos (2\pi{\Delta}/{\lambda})}{3+\cos (2\pi{\Delta}/{\lambda})}\right),
\end{equation} 
where $e_{max}$ is the maximum efficiency of the unit -- in theory this is 1, however we find a value slightly less than this sometimes fits the data better. The term $\Delta$ is the optical path length of the FLC, and according to \citet{gisler03} is given by \begin{equation} \label{eqn:delta}\Delta=\frac{\lambda_0}{2}+Cd\left ( \frac{1}{\lambda^2}-\frac{1}{\lambda_{0}^{2}} \right ),\end{equation} where $\lambda_0$ is the wavelength of peak efficiency (i.e. the half-wave wavelength) and the terms $C$ -- describing the birefringence of the crystal -- and $d$ -- the layer thickness -- can be treated as a single term. Thus the modulation efficiency can be determined as a function of wavelength by fitting $e_{max}$, $\lambda_0$, $Cd$ and the blackbody temperature of the source, $T_{bb}$.

\begin{figure}
\includegraphics[width=8.5cm]{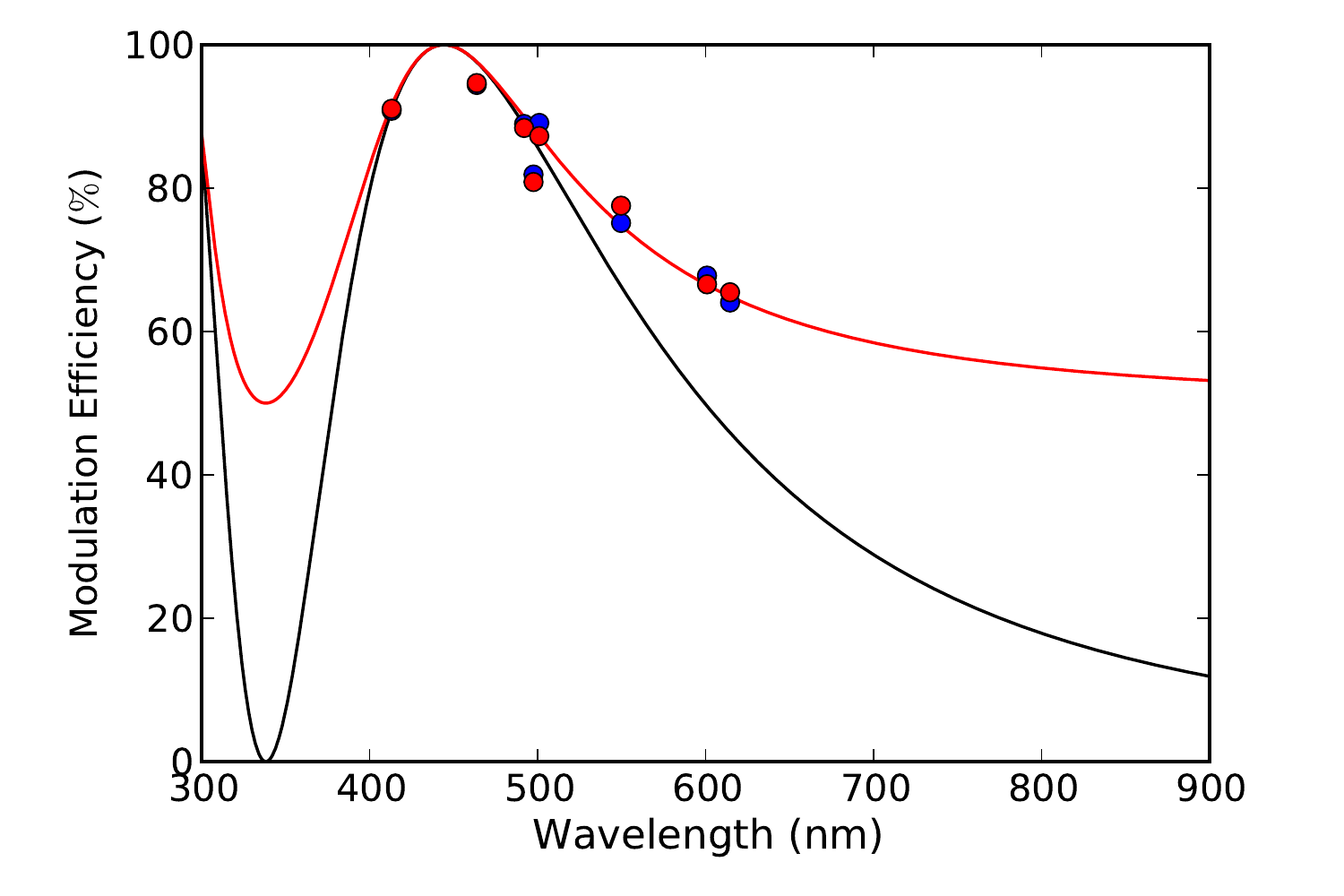}
\caption{Shown is the laboratory data (blue dots) taken to calibrate the Meadowlark modulator. The red line shows the (high polarization) modulation efficiency curve that best fits the data, with the red points corresponding to the exact bandpass of the data points. The black line shows the low polarization approximation modulation curve for the same fit parameters. The points are shown to correspond to their effective wavelength, and are left to right: 400NB, 500SP, \sdssg{} , 425SP, 500NB, Clear, 600NB, \sdssr. Although the bluer detectors were used, the 425SP effective wavelength is longer than typical owing to the extreme redness of the source (2551$\pm$149 K blackbody).}
 \label{fig:ML}
\end{figure}

Figure \ref{fig:ML} shows the fit obtained to the calibration data for the ML modulator; the fit parameters are given in Table \ref{tab:mod}. Also shown is the low polarization approximation for $e(\lambda)$. Astronomical observations made with HIPPI-2 are almost exclusively of objects for which the low polarization approximation (see appendix \ref{sec:apa}) is appropriate (up to 10\%). In this case the modulation efficiency is given by 

\begin{equation}\label{eq:e_lp} e(\lambda)= e_{max} \left(\frac{1-\cos (2\pi{\Delta}/{\lambda})}{2} \right). 
\end{equation}

The BNS modulator was originally calibrated for HIPPI in early 2014 in the laboratory in a similar way to the ML unit. However, it has become apparent that its performance has changed over time with the $\lambda_0$ value in equation \ref{eqn:delta} shifting to longer wavelengths and this performance drift has accelerated. Consequently we have since employed a different method to calibrate the modulators, where multi-band observations of high polarization standards are used as known sources in the calibration. In this case it is equation \ref{eq:e_lp} that is used rather than equation \ref{eq:e_hp} in the bandpass model to fit the data, but otherwise the procedure is the same.

Table \ref{tab:hp_std} gives details of the high polarization standard stars we have employed to calibrate modulator performance. Table \ref{tab:mod} shows the parameters fit for the ML and BNS modulators used with HIPPI-2, as well as the Micron Technologies (MT) modulator previously used with HIPPI and Mini-HIPPI. The ML modulator has a bluer $\lambda_0$ than either of the other two. The BNS modulator has been used a lot; we have broken its usage down into seven eras, the last five of which correspond to HIPPI-2 runs. Clearly $\lambda_0$ has increased over time -- by $\sim$100 nm. The modulator was used frequently in 2018, with over 50 nights of observing, but it isn't clear what the cause of the performance drift is. By contrast the MT modulator's performance is unchanged despite 5 years use.

\begin{table*}
\caption{Modulator Parameters}
\begin{tabular}{lclllll}
\hline
\hline
Modulator & Era 	& Instrument 	& Data from 			& \multicolumn{1}{c}{$\lambda_0$}	& \multicolumn{1}{c}{$Cd$}  	& $e_{max}$			\\
          &         &               &                       & \multicolumn{1}{c}{(nm)}      &   \multicolumn{1}{c}{($\times10^7$ nm$^3$)}    &   \\
\hline
\hline
ML  & 		& HIPPI-2		& 2018 Laboratory		& 444.2$\pm$ 2.7 & 3.163$\pm$0.476		& 1.000$\pm$0.021	\\
ML  & 1     & HIPPI-2       & 2018 to 2019          & 455.2$\pm$ 1.9 & 2.677$\pm$0.103      & 1.000             \\
\hline
BNS & 		& HIPPI			& 2014 Laboratory		& 504.6$\pm$ 2.4 & 2.277$\pm$0.175		& 0.977$\pm$0.009	\\
BNS & 1		& HIPPI			& 2014 to 2015			& 494.8$\pm$ 1.6 & 1.738$\pm$0.060		& 0.977				\\
BNS & 2		& HIPPI			& 2016 to 2017			& 506.3$\pm$ 2.9 & 1.758$\pm$0.116		& 0.977				\\
BNS & 3		& HIPPI-2		& 2018 Jan -- May		& 512.9$\pm$ 3.9 & 2.367$\pm$0.177		& 0.977				\\
BNS & 4		& HIPPI-2		& 2018 Jul				& 517.5$\pm$16.1 & 2.297$\pm$0.924		& 0.977				\\
BNS & 5		& HIPPI-2		& 2018 Aug 16 -- 23		& 546.8$\pm$ 6.0 & 2.213$\pm$0.261		& 0.977				\\
BNS & 6		& HIPPI-2		& 2018 Aug 24 -- 27		& 562.7$\pm$ 4.7 & 2.329$\pm$0.192		& 0.977				\\
BNS & 7		& HIPPI-2		& 2018 Aug 29 -- Sep 2	& 595.4$\pm$ 4.8 & 1.615$\pm$0.145		& 0.977				\\
\hline
MT$^a $ & 	& HIPPI			& 2014 Laboratory		& 505.\0$\pm$ 5.  & 1.75\0$\pm$0.05 		& 0.98 $\pm$0.00	\\
MT  & 		& HIPPI/MHIPPI	& 2014 to 2018			& 507.6$\pm$ 2.6 & 1.837$\pm$0.128		& 0.980				\\
\hline
\hline
\end{tabular}
\begin{flushleft}
Notes: \\
Errors given for parameters fit. \\
$^a$ From \citet{bailey15}, given to fewer decimal places. \\
\end{flushleft}
\label{tab:mod}
\end{table*}

\begin{table*}
\caption{Polarized Standard Stars}
\begin{tabular}{lrllllllrll}
\hline
Standard	&	m$_V$	&   SpT	& 	E(B$-$V) &	R$_V$	& p$_{max}$ 	& $\lambda_{max}$  &  K	& 	PA$^a$  & References	& Desig.\\
            &           &       &            &          & \multicolumn{1}{c}{(\%)}          & \multicolumn{1}{c}{($\mu$m)}         &   & \multicolumn{1}{c}{($^\circ$)} &              &       \\ 
\hline
HD  23512   &   8.09	&	A0  &   0.36	&	3.3		&	3.2		&	0.61	&	1.02	&	 30.0	& 1, 2, 3. & A \\
HD  80558	&	5.93	&	B6	&	0.60	&	3.35	&	3.34	&	0.597	&	1.33	&	163.3	& 4. & B \\
HD  84810   &	3.40	& 	F8	&	0.34	& 	3.1 	& 	1.62	& 	0.57	& 	1.15	& 	100.0	& 2, 4. & C \\
HD 111613 &	5.72	&	A1	&	0.39	&	3.1$^b$	&	3.2 	&	0.56	&	0.94	&	 81.0	& 1. & D \\
HD 147084	&	4.57	& 	A4	&	0.72	& 	3.9 	& 	4.34	& 	0.67	& 	1.15	& 	 32.0 	& 5, 6. & E \\
HD 149757	&	2.56	&	O9	&	0.32	&	3.09	&	1.48	&	0.598	&	1.00	&	127.4	& 7, 8. & F \\
HD 154445	& 	5.61	&	B1	& 	0.42	& 	3.15	& 	3.73	& 	0.558	& 	0.95	& 	 90.1 	& 2, 4, 6. & G \\
HD 160529	& 	6.66	&	A2	& 	1.29	& 	3.1 	& 	7.31	& 	0.543	& 	1.15	& 	 20.4 	& 2, 4.	& H \\
HD 161056	&	6.32	&	B1.5 &	0.59	&	3.13	&	4.02	&	0.572	&	1.43	&	 67.5	& 9. & I \\
HD 187929	& 	3.80	&	F6	&	0.18	& 	3.1 	& 	1.76	& 	0.56	& 	1.15	& 	 93.8	& 4. & J \\
HD 203532	&	6.38	&	B3	&	0.28	&	3.37	&	1.23	&	0.574	&	1.39	&	127.8	& 9. & K \\
HD 210121	&	7.68	&	B7	&	0.31	&	2.42	&	1.38	&	0.434	&	0.55	&	155.4	& 9. & L \\
\hline
\end{tabular}
\begin{flushleft}
Notes: \\
References: (1) \citet{serkowski74}, (2) \citet{hsu82}, (3) \citet{guthrie87}, (4) \citet{serkowski75}, (5) \citet{wilking80}, (6) \citet{martin99}, (7) \citet{mcdavid00}, (8) \citet{patriarchi01}, (9) \citet{bagnulo17}. \\
$^a$ PA chosen to reflect that expected in the \sdssg{}  filter. \\
$^b$ The value of R$_V$ is assumed, and HD 111613 has been used to calibrate PA, but not modulator performance. \\
\end{flushleft}
\label{tab:hp_std}
\end{table*}

\subsubsection{PA Correction}
\label{sec:pa_corr}

During a run, observations made of polarized standard stars (table \ref{tab:hp_std}) in either the \sdssg{}  or Clear filters are used to determine the position angle alignment of the instrument. The average difference between the PA values from the literature and the PA from our measurements, denoted $\theta$, is determined and all the data rotated according to \begin{equation}q=q_i\cos \theta+ u_i\sin \theta\end{equation}\begin{equation}u=u_i\cos \theta- q_i\sin \theta\end{equation} where the $i$ subscript denotes the instrument reference frame. The precision of the literature measurements is not much better than a degree, yet typically the standard deviation of PA measurements made with HIPPI-2 is 0.5$^\circ$ or better.

Observations of polarized standards in other filters are made to check for wavelength dependent effects. The PA of polarized standards can change slightly with wavelength, but any deviation, $\Delta\theta$, much larger than a degree is considered to be an instrumental effect. 

The majority of the time there is no significant rotation with wavelength. However, during the 2018JUL and 2018AUG runs\footnote{See Section \ref{sec:performance} for a full description of observing runs.} a significant $\Delta\theta$ was detected at short wavelengths, which we infer is associated with the performance drift of the BNS modulator. For 2018AUG $\Delta\theta$ was greatest for the 500SP filter, 5.8$^\circ$, reducing to 2.6$^\circ$ for the 425SP filter. The effect was similar for 2018JUL: 5.6$^\circ$ for 500SP, and 3.3$^\circ$ for 425SP. Observations made in these bands are counter-rotated by a corresponding amount as a correction. For observations made in Clear with a $\lambda_{eff}$ less than the mean of the \sdssg{}  polarized standards, a correction was calculated by fitting a parabola to the $\Delta\lambda$ and $\lambda_{eff}$ values of the \sdssg{}, 500SP and 425SP filters to get a function for $\Delta\theta(\lambda_{eff})$. Small corrections were also applied to 2018FEB and 2018MAR data using a similar method: 2.7$^\circ$ at 425SP and 1.35$^\circ$ at 500SP.


Similar corrections for the ML modulator have not been required for 425SP or 500SP bands, but a correction of $\Delta\theta = -$14.65$^\circ$ to the U band data from the 2019MAR run was required.

\subsubsection{TP Correction}
\label{sec:tp}

The last correction applied is that for the TP\footnote{When the TP is large compared to the polarized standard polarization magnitudes, the order of the TP and PA corrections need to be swapped. When the TP is small however, performing the PA correction first has the benefit of determining the TP in the sky reference frame. TP determinations can then be combined easily from back-to-back runs where the instrument is mounted at a different PA}. This is the zero-point correction, or the polarization we measure when observing an unpolarized source. The telescope optics impart a small polarization on every measurement recorded. On an equatorial telescope we can treat this polarization as a constant offset. While the telescope is the main cause of this zero point offset, it is possible that when the actual telescope polarization is small there may be significant contributions to the zero point from instrumental sources as well.  For each filter, detector and aperture combination we calculate a TP in terms of $q$ and $u$ based on the average of low polarization standard stars observed. Table \ref{tab:lp_std} gives a list of the standards we have employed. The list has been kept short deliberately with the aim of collecting enough comparable data on the standards to eventually determine the offsets between them -- and better identify undesirable variability -- but at present each is assumed to be zero to determine the TP. 

Although there are polarization values given for these stars in the literature, they either come from PlanetPol observations (those from \citealp{bailey10}) where the bandpass was quite different to the typical HIPPI/-2 observation, or they are from other observations we have made with this same method of determining TP. However, all of the low polarization stars we use are near enough to the Sun so that interstellar polarization will be very low \citep{cotton16,cotton17b} and have spectral types not associated with intrinsic polarization \citep{cotton16,cotton16b}. The furthest standards reside in a part of the northern sky found to have an interstellar polarization per distance about an order of magnitude less than is common in the southern sky \citep{bailey10,cotton16,cotton17b}.

In the case that an observation is made and no specific standards have been observed with the same exact set-up, the combination with the same filter and detector and closest aperture size is used first. If this fails, the combination with the closest effective wavelength to the target is used instead.

\begin{table*}
\caption{Low Polarization Standard Stars}
\centering
\begin{tabular}{llrlrrrll}
\hline
Standard	&  Hem$^a$	&	m$_V$	&	SpT 	&	\multicolumn{1}{c}{\0d}	& \multicolumn{1}{c}{\0q}     & \multicolumn{1}{c}{\0u}	    & References & Desig.	\\
            &           &           &           & \multicolumn{1}{c}{\0(pc)}  & \multicolumn{1}{c}{\0(ppm)} & \multicolumn{1}{c}{\0(ppm)} &           &       \\
\hline
HD   2151   &  S	&	2.79	&	G0V  	&  7.5	& $-$8.6$\pm$2.5	& $-$1.6$\pm$2.5	& \citet{cotton16} & A\\
HD  10700	&  S	&	3.50	&	G8V		&  3.7  & 1.3$\pm$3.1	& 0.3$\pm$3.0	& \citet{cotton17b} & B\\
HD  49815   &  S	&  $-$1.46	&	A1V+DA	&  2.6 	& $-$3.7$\pm$1.7	& $-$4.0$\pm$1.7	& \citet{cotton16} & C\\
HD 102647   &  N,S	&	2.13	&	A3Va	& 11.0	& 0.8$\pm$1.1	& 2.2$\pm$0.8	& \citet{bailey10} & D\\
HD 102870	&  S	&	3.60	&	F9V		& 11.1	& 3.3$\pm$1.4	& $-$0.1$\pm$1.4	& \citet{bailey10} & E\\
HD 127762	&  N	&	3.02	&	A7IV	& 26.6	& $-$2.8$\pm$1.6	& $-$2.2$\pm$1.6	& \citet{bailey10} & F\\
HD 128620J	&  S$^b$ & $-$0.10		&	G2V+K1V &  1.3	& 5.7$\pm$1.9	& 14.4$\pm$1.9	& \citet{bailey17} & G\\
HD 140573	&  N,S	&	2.63	&	K2IIIb 	& 25.4	& $-$2.3$\pm$2.9	& 3.9$\pm$1.0	& \citet{bailey10} & H\\
\hline
\end{tabular}
\begin{flushleft}
Notes: \\
$^a$ Indicates the hemisphere(s) in which the standard has been used. \\
$^b$ HD 128620J ($\alpha$ Cen) is used predominantly on small telescopes where the night-to-night precision is greater than the reported polarization. \\
\end{flushleft}
\label{tab:lp_std}
\end{table*}

\section{Instrument Performance}
\label{sec:performance}

The performance of HIPPI-2 has been evaluated based on observations obtained during 2018 and early 2019 on three telescopes. Observations with the WSU 60 cm telescope were obtained on 2018 Jan 23, May 4--5 and 9--11 and 2019 Feb 11-15. Observations with the 3.9 m AAT were obtained on 2018 Feb 1--5, Mar 23--Apr 8, Jun 10--25, Aug 16--Sep 2 and 2019 Mar 15 to 26. Observations with the Gemini North telescope were obtained in Director's Discretionary time on 2018 Jul 4--6. Table \ref{tab:runs} lists the telescope and instrument configurations for each run. In the following discussion we refer to the individual runs using the names given in the first column of table \ref{tab:runs}. 

\begin{table*}
\caption{Summary of Runs for HIPPI-2}
\centering
\begin{tabular}{lllllll}
\hline
Run         &S/R$^a$&   \multicolumn{1}{c}{Run Duration}       &   \0Tel.        &   \multicolumn{1}{c}{Focus}      &   \multicolumn{1}{c}{Mod.}        &   Comments \\
            &       &   \multicolumn{1}{c}{(UT)}                &               &   \multicolumn{1}{c}{(f/)}                           &                      &           \\
\hline
2018JAN     &   & 2018-01-23                &   WSU         &   10.5*   &   BNS-E3      &   WSU Commissioning Run.\\
2018FEB     & A & 2018-02-02                &   AAT         &   15      &   BNS-E3      &   AAT Commissioning Run.\\
            & B & 2018-02-03                &               &           &               &   Back-end adjustment.\\
        & C$^b$ & 2018-02-04                &               &           &               &   Back-end adjustment.\\
        & D$^b$ & 2018-02-04 to 2018-02-05  &               &           &               &   Alignment adjusted.\\
2018MAR     &   & 2018-03-23 to 2018-04-07  &   AAT         &   \08*    &   BNS-E3      &   Back-end redesigned.\\
2018MAY     &   & 2018-05-04 to 2018-05-11  &   WSU         &   10.5*   &   BNS-E3      &   \\
2018JUN     &   & 2018-07-04 to 2018-07-06  &   Gemini Nth  &   16      &   ML-E1       &   Clone instrument.$^d$ \\
2018JUL$^c$ &   & 2018-07-10 to 2018-07-25  &   AAT         &   \08*    &   BNS-E4      &   Rapid modulator evolution.\\
2018AUG$^c$ &   & 2018-08-16 to 2018-08-23  &   AAT         &   \08*    &   BNS-E5      &   Rapid modulator evolution.\\
            &   & 2018-08-24 to 2018-08-27  &               &           &   BNS-E6      &   Rapid modulator evolution.\\
            &   & 2018-08-29 to 2018-09-02  &               &           &   BNS-E7      &   Rapid modulator evolution.\\
2019FEB     &   & 2019-02-11 to 2019-02-15  &   WSU         &   10.5*   &   ML-E1       &   \\
2019MAR     &   & 2019-03-15 to 2019-03-26  &   AAT         &   15      &   ML-E1       &   650LP replaced with U.\\
\hline
\end{tabular}
\begin{flushleft}
Notes: \\
* Indicates a focal configuration requiring the use of the negative achromatic lens -- the effective focal ratio is f/ twice the number given. The two different focal arrangements on the AAT use different secondary mirrors. \\
$^a$ S/R indicates a sub-run, i.e. where the instrument has been removed from the telescope mid-run and then remounted. Ordinarily this operation requires a new PA calibration, but allows TP measurements to be combined. However, for the 2018FEB-B and 2018FEB-C runs the instrument was altered compared to the previous sub-run and new TP calibrations were acquired. \\
$^b$ and $^c$ indicate that the TP has been combined between these runs or sub-runs, this is possible where the instrument and telescope performance is stable.\\
$^d$ The clone is a complete copy of the original instrument. The aperture wheel is 3D printed and varies between units; nominal aperture sizes have been assumed for the 2018JUN run. The clone instrument used a different pair of blue PMT units for the 2018JUN run than have otherwise been used with HIPPI-2; these PMTs were used for early HIPPI runs.
\end{flushleft}
\label{tab:runs}
\end{table*}

\subsection{Throughput}

HIPPI-2 improves on the optical throughput of HIPPI through the use of a simpler optical system and the use of more efficient filters. Using the bandpass model described in section \ref{sec:bandpass} we find that the expected improvement in instrument throughput amounts to about 20\% in Clear and about 65\% in the \sdssg{} filter. Inspection of the measured intensity in actual AAT observations indicates the real improvement is a little better than these figures predict. Additional gains probably come from the improved telescope throughput due to the use of the f/15 secondary which has a smaller central obstruction than the f/8 configuration used with HIPPI, as well as from the ability to use larger apertures with HIPPI-2 that eliminate any spillage of light due to seeing.  

\subsection{Telescope Polarization}

Measurements of the zero point correction (or TP) were made by observing low polarization standard stars as described in section \ref{sec:tp}. Results for the the equatorially mounted telescopes (AAT and WSU) are listed in tables \ref{tab:tp_wsu} and \ref{tab:tp_aat}. A unique TP determination was made for each filter and aperture combination used. For each telescope and each run set the individual TP determinations are listed in order of effective wavelength. For most measurements the TP magnitude is greatest in the bluest wavelength bands. The TP PA is very similar between bands most of the time, but does appear to rotate slightly away from the mean in the bluest bands -- the low TP in 2018JUL/AUG and 2019MAR accentuates this rotation. 

As noted in section \ref{sec:tp} while the telescope polarization is the main contributor to the zero point polarization measured with HIPPI-2 there are likely to be small contributions from residual instrumental effects as well. One example of this is that there are minor differences between TP measured in the same band but with different apertures. In part this may be due to using different standard stars. However, we believe there are other significant factors. During the 2018MAR run we used different centering strategies for the different aperture sizes. In the two smallest apertures the standards were re-centered at each PA; in the larger apertures centering was performed only at $\textrm{PA}=0^{\circ}$. This may lead to small zero point offsets due to the centering effects described in section \ref{sec:limits}

\begin{table*}
\caption{Telescope Polarization by Run at WSU with HIPPI-2.}
\centering
\tabcolsep 3 pt
\begin{tabular}{lccrr|cccccccc|rr}
\hline
\hline
Run$^a$ &   Fil  & PMT  & \multicolumn{1}{c}{Ap} & $\lambda_{eff}$ &   \multicolumn{8}{c|}{Standard Observations} & \multicolumn{1}{c}{$p \pm \Delta p$} & \multicolumn{1}{c}{$PA \pm \Delta PA$}\\
& & &\multicolumn{1}{c}{(\arcsec)}&(nm)& A & B & C & D & E & F & G & H &\multicolumn{1}{c}{(ppm)} & \multicolumn{1}{c}{($^{\circ}$)} \\
\hline
\hline
2018JAN & Clear    & B & 58.9 & 469.6 & 0 & 0 & 3 & 0 & 0 & 0 & 0 & 0 & 41.4 $\pm$ 2.6 & 128.3 $\pm$ 1.8 \\
\hline
2018MAY & \sdssg{} & B & 58.9 & 464.9 & 0 & 0 & 4 & 0 & 0 & 0 & 0 & 0 & 25.4 $\pm$ 2.6 & 82.8 $\pm$ 2.9 \\
2018MAY & Clear    & B & 58.9 & 473.4 & 0 & 0 & 6 & 5 & 0 & 0 & 0 & 0 & 27.7 $\pm$ 2.0 & 92.8 $\pm$ 2.0 \\
2018MAY & Clear    & R & 58.9 & 601.9 & 0 & 0 & 0 & 0 & 0 & 1 & 0 & 0 & 33.7 $\pm$ 3.9 & 79.1 $\pm$ 3.2 \\
\hline
2019FEB & \sdssg{} & B & 58.9 & 463.1 & 0 & 0 & 4 & 0 & 0 & 0 & 0 & 0 & 23.9 $\pm$ 1.7 & 28.9 $\pm$ 2.0 \\
2019FEB & Clear    & B & 58.9 & 467.3 & 0 & 0 & 3 & 0 & 0 & 0 & 0 & 0 & 12.1 $\pm$ 3.1 & 47.1 $\pm$ 7.6 \\
\hline
\hline
\end{tabular}
\begin{flushleft}
Notes: \\
The key for the letters denoting the low polarization standards is in table \ref{tab:lp_std}. \\
\end{flushleft}
\label{tab:tp_wsu}
\end{table*}

The TP recorded on the WSU telescope has always been very low -- between about 10 to 40 ppm. This compares favourably to the UNSW telescope where the TP has been around 60 to 90 ppm \citep{bailey17,bailey19}. The small differences between runs might be down to refinements we have made to the way the instrument is mounted over time, or it could be related to the dust pattern on the mirrors. Regardless, there have been no significant shifts during a run. 

\begin{table*}
\caption{Telescope Polarization by Run at the AAT with HIPPI-2.}
\centering
\tabcolsep 3 pt
\begin{tabular}{lccrr|cccccccc|rr}
\hline
\hline
Run &   Fil  & PMT  & \multicolumn{1}{c}{Ap} & $\lambda_{eff}$ &   \multicolumn{8}{c|}{Standard Observations} & \multicolumn{1}{c}{$p \pm \Delta p$} & \multicolumn{1}{c}{$PA \pm \Delta PA$}\\
&&&\multicolumn{1}{c}{(\arcsec)}&(nm)& A & B & C & D & E & F & G & H &\multicolumn{1}{c}{(ppm)} & \multicolumn{1}{c}{($^{\circ}$)} \\
\hline
\hline
2018FEB-A & \sdssg{}& B & 16.8 & 464.3 & 0 & 0 & 0 & 0 & 1 & 0 & 0 & 0 & 225.4 $\pm$ 3.8 & 88.1 $\pm$ 0.5 \\
\hline
2018FEB-B & 425SP   & B & 16.8 & 399.0 & 0 & 0 & 1 & 0 & 0 & 0 & 0 & 0 & 303.6 $\pm$ 3.0 & 91.4 $\pm$ 0.3 \\
2018FEB-B & \sdssg{}& B & 16.8 & 463.5 & 0 & 0 & 2 & 1 & 0 & 0 & 0 & 0 & 192.6 $\pm$ 1.1 & 88.1 $\pm$ 0.2 \\
2018FEB-B & Clear   & B & 16.8 & 467.0 & 0 & 0 & 2 & 1 & 0 & 0 & 0 & 0 & 190.5 $\pm$ 1.0 & 88.7 $\pm$ 0.2 \\
2018FEB-B & Clear   & B &  9.2 & 467.2 & 0 & 0 & 2 & 1 & 0 & 0 & 0 & 0 & 186.4 $\pm$ 1.1 & 86.8 $\pm$ 0.2 \\
2018FEB-B & \sdssr{}& B & 16.8 & 602.6 & 0 & 0 & 1 & 0 & 0 & 0 & 0 & 0 & 129.6 $\pm$ 2.0 & 88.1 $\pm$ 0.4 \\
\hline
2018FEB-C/D & 500SP & B & 16.8 & 434.7 & 0 & 0 & 1 & 0 & 0 & 0 & 0 & 0 & 215.6 $\pm$ 1.0 & 90.0 $\pm$ 0.1 \\
2018FEB-C/D &\sdssg{}&B & 16.8 & 463.6 & 0 & 0 & 5 & 3 & 0 & 0 & 0 & 0 & 178.9 $\pm$ 0.7 & 87.7 $\pm$ 0.1 \\
2018FEB-C/D & Clear & B & 16.8 & 469.5 & 0 & 0 & 5 & 3 & 1 & 0 & 0 & 0 & 179.7 $\pm$ 0.8 & 87.7 $\pm$ 0.1 \\
2018FEB-C & \sdssg{}& R & 16.8 & 481.8 & 0 & 0 & 1 & 0 & 0 & 0 & 0 & 0 & 168.7 $\pm$ 0.9 & 81.2 $\pm$ 0.2 \\
2018FEB-C & \sdssr{}& R & 16.8 & 622.2 & 0 & 0 & 1 & 0 & 0 & 0 & 0 & 0 & 109.0 $\pm$ 1.2 & 86.3 $\pm$ 0.3 \\
2018FEB-C & 650LP   & R & 16.8 & 720.7 & 0 & 0 & 1 & 0 & 0 & 0 & 0 & 0 &  81.2 $\pm$ 1.9 & 90.3 $\pm$ 0.7 \\
\hline
2018MAR & 425SP     & B & 15.7 & 403.0 & 0 & 0 & 3 & 3 & 3 & 0 & 0 & 0 & 183.1 $\pm$ 2.8 &  4.0 $\pm$ 0.4 \\
2018MAR & 500SP     & B & 15.7 & 440.9 & 0 & 0 & 3 & 3 & 3 & 0 & 0 & 0 & 145.5 $\pm$ 1.2 &  4.0 $\pm$ 0.2 \\
2018MAR & \sdssg{}  & B & 15.7 & 466.3 & 0 & 0 & 3 & 3 & 3 & 0 & 0 & 0 & 130.0 $\pm$ 0.9 &  0.9 $\pm$ 0.2 \\
2018MAR & Clear     & B &  8.6 & 471.3 & 0 & 0 & 2 & 3 & 0 & 0 & 0 & 0 & 114.8 $\pm$ 0.7 &178.7 $\pm$ 0.2 \\
2018MAR & Clear     & B &  5.3 & 472.9 & 0 & 0 & 0 & 3 & 0 & 0 & 0 & 0 & 125.1 $\pm$ 1.3 &177.2 $\pm$ 0.3 \\
2018MAR & Clear     & B & 15.7 & 485.1 & 0 & 0 & 4 & 4 & 3 & 0 & 0 & 5 & 130.8 $\pm$ 0.7 &  1.2 $\pm$ 0.1 \\
2018MAR & V         & B & 15.7 & 533.2 & 0 & 0 & 3 & 0 & 0 & 0 & 0 & 0 & 125.6 $\pm$ 0.8 &  2.5 $\pm$ 0.2 \\
2018MAR & \sdssr{}  & R & 15.7 & 623.3 & 0 & 0 & 2 & 3 & 2 & 0 & 0 & 0 & 113.6 $\pm$ 1.4 &  1.8 $\pm$ 0.4 \\
2018MAR & 650LP     & R & 15.7 & 722.3 & 0 & 0 & 2 & 3 & 2 & 0 & 0 & 0 & 107.8 $\pm$ 1.9 &  2.8 $\pm$ 0.5 \\
\hline
2018JUL/AUG &425SP  & B & 11.9 & 407.3 & 3 & 2 & 2 & 0 & 1 & 0 & 0 & 3 & 19.8 $\pm$ 6.2 & 49.7 $\pm$ 9.4 \\
2018JUL/AUG &500SP  & B & 11.9 & 445.0 & 2 & 2 & 2 & 2 & 1 & 0 & 0 & 2 & 18.6 $\pm$ 1.4 & 41.2 $\pm$ 2.2 \\
2018JUL/AUG&\sdssg{}& B & 11.9 & 470.4 & 3 & 3 & 2 & 2 & 1 & 0 & 0 & 2 & 13.6 $\pm$ 1.1 & 80.9 $\pm$ 2.2 \\
2018JUL/AUG& Clear  & B & 11.9 & 489.4 & 3 & 3 & 2 & 2 & 2 & 0 & 0 & 2 & 10.6 $\pm$ 0.9 & 79.6 $\pm$ 2.6 \\
2018JUL/AUG& V      & B & 11.9 & 537.9 & 1 & 1 & 2 & 0 & 1 & 0 & 0 & 2 & 20.7 $\pm$ 1.5 & 87.0 $\pm$ 2.1 \\
2018JUL/AUG&\sdssr{}& B & 11.9 & 605.4 & 1 & 2 & 2 & 0 & 1 & 0 & 0 & 3 & 18.6 $\pm$ 1.4 & 81.3 $\pm$ 2.8 \\
2018JUL/AUG&\sdssr{}& R & 11.9 & 625.6 & 4 & 2 & 0 & 2 & 2 & 0 & 0 & 1 & 12.5 $\pm$ 1.2 & 88.5 $\pm$ 2.7 \\
2018JUL/AUG& 650LP  & R & 11.9 & 725.7 & 3 & 2 & 0 & 2 & 2 & 0 & 0 & 1 &  8.1 $\pm$ 1.9 & 75.8 $\pm$ 7.0 \\
\hline
2019MAR & U         & B & 12.7 & 380.6 & 0 & 0 & 2 & 2 & 0 & 0 & 0 & 1 & 103.7 $\pm$ 7.6 & 88.4 $\pm$ 2.1 \\
2019MAR & 425SP     & B & 12.7 & 398.3 & 0 & 0 & 2 & 0 & 0 & 0 & 0 & 0 &   4.7 $\pm$ 1.1 & 56.1 $\pm$ 7.1 \\
2019MAR & 500SP     & B & 12.7 & 434.4 & 0 & 0 & 1 & 1 & 0 & 0 & 0 & 0 &  67.3 $\pm$ 2.1 &110.0 $\pm$ 0.9 \\
2019MAR & \sdssg{}  & B & 12.7 & 462.9 & 0 & 0 & 3 & 2 & 0 & 0 & 0 & 0 &   9.5 $\pm$ 0.8 & 79.3 $\pm$ 2.1 \\
2019MAR & Clear     & B & 12.7 & 464.0 & 0 & 0 & 3 & 0 & 0 & 0 & 0 & 0 &   9.7 $\pm$ 1.1 & 36.3 $\pm$ 3.1 \\
2019MAR & V         & B & 12.7 & 540.5 & 0 & 0 & 0 & 0 & 0 & 0 & 0 & 1 &  21.3 $\pm$ 6.5 & 37.5 $\pm$ 9.1 \\
2019MAR & \sdssr{}  & B & 12.7 & 602.6 & 0 & 0 & 1 & 0 & 0 & 0 & 0 & 0 &  13.1 $\pm$ 5.6 &105.1 $\pm$14.1 \\
\hline
\hline
\end{tabular}
\begin{flushleft}
Notes: \\
The key for the letters denoting the low polarisation standards is in table \ref{tab:lp_std}. \\
During each run, one aperture setting is chosen as a default, with which most observations are made. We have omitted from this table TP determinations made in other apertures where only a single observation was made.\\
\end{flushleft}
\label{tab:tp_aat}
\end{table*}

\begin{figure}
\includegraphics[width=\columnwidth]{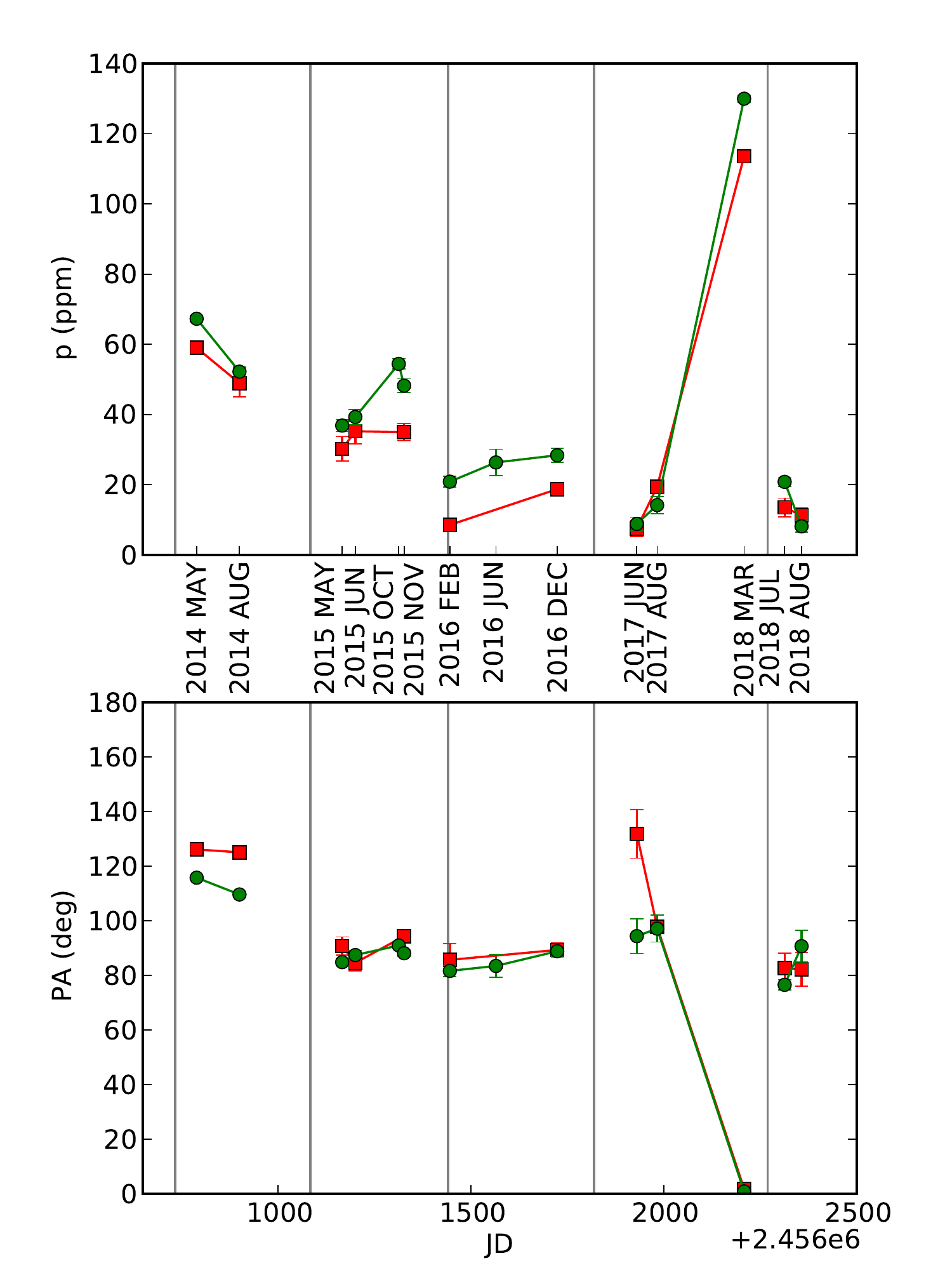}
\caption{Telescope polarization (TP) at the AAT f/8 focus in two bands: \sdssg{} (green circles) and \sdssr{} (red squares) plotted against time (JD). The upper panel shows the magnitude of the polarization, the lower panel shows the position angle. The run designations are given between the two panels. The vertical grey lines show when the primary mirror was realuminised. The HIPPI data and some of the HIPPI-2 data shown here and/or reported in table \ref{tab:tp_aat} has been previously reported \citep{bailey15, cotton16, marshall16, bott16, cotton17a, cotton17b, bott18, cotton19a, cotton19b, bailey19}, but the data have been reprocessed to benefit from refinements in the software.}
 \label{fig:AATf8TP}
\end{figure}

In figure \ref{fig:AATf8TP} are plotted all the TP measurements made with HIPPI and HIPPI-2 at the f/8 focus of the AAT in both \sdssg{} and \sdssr{} bands (which are the most consistently observed filter bands). The grey vertical lines represent realuminisation of the primary mirror. A number of observations can be made. The magnitude of the TP is usually lower in \sdssg{} than \sdssr{}; this is consistent with what we see in table \ref{tab:tp_aat}. The magnitude of the TP was reduced at every realuminisation from 2014 to 2017 where it reached around 10 ppm. The magnitude of the TP tends to increase with time following realuminisation -- this can reasonably be ascribed to the inevitable buildup of dust on the main mirror with time. The TP PA has been fairly consistent, with the exception of the 2018MAR run, which is probably reflective of the contribution to the TP of the secondary mirror. The 2018MAR result can most easily be explained by the primary mirror being marked prior to the run\footnote{Possibly on March 7th when the f/15 secondary was removed for realuminising. This coating was of poor quality, which is why f/8 was used for 2018MAR.} since the TP returns to a normal level following realuminising. If instrumental polarization was contributing more then we would expect the PA in 2017 to be different to all the 2018 runs corresponding to the change from HIPPI to HIPPI-2.

The TP was also very high during the 2018FEB run. The PA is 90$^\circ$ different to the similarly high 2018MAR run, suggesting a different cause. We ascribe this to the condition of the f/15 secondary mirror, which is rarely used. The AAT primary mirror is usually realuminised every year. The f/8 secondary mirror was last realuminised in 2004, and prior to the 2019MAR run it had been more than 20 years since the f/15 secondary was realuminised (S. Lee, priv. comm.). While the f/8 secondary is always well protected from falling dust, the f/15 secondary shares a mounting with the f/36 secondary and has occasionally been in an upward facing position without the side dust covers installed. The f/15 secondary was realuminised in March 2018 and again in September of 2018. This explains why the TP is much lower in the 2019MAR run than it was for the earlier runs. 

\subsection{Polarization Precision}
\label{sec:precision}

The polarization precision achievable with HIPPI-2 has been evaluated by making repeat observations of bright low-polarization stars in the same way as the analysis of HIPPI presented by \citet{bailey15}. Amongst the stars used for this analysis are many of our low polarization standards, as well as a number of other stars with small polarizations unlikely to be variable. Table \ref{tab:prec_aat} shows such measurements made at the AAT during the 2018MAR run using a 15.7\arcsec{} aperture; the observations are grouped by filter band. In the table alongside the error weighted mean of the stokes $q$ and $u$ values are the associated standard deviations, $\sigma$, and the average internal error of the individual measurements, $\delta$.

The values of $\sigma$ are a conservative estimate of the precision we are achieving. However, they will tend to underestimate our ultimate precision as they include a contribution from the internal statistical error of each measurement. To attempt to allow for this we also calculate what we refer to as the error variance, calculated as: \begin{equation}e=\left\{\begin{matrix}\sqrt{\sigma^2-\delta^2} & \sigma>\delta \\ 0 & \sigma\leq\delta \end{matrix}\right. \end{equation} We use the subscript $p$ to denote the mean of $q$ and $u$ determinations of $\sigma$ and $e$. $e_p$ is thus an estimate of the precision we would expect to see in repeat observations if the internal errors were very small. This metric deals poorly with individual instances where $\sigma<\delta$, and is thus most useful only when examining the mean of many measurements. Based on the $e_p$ values in table \ref{tab:prec_aat} HIPPI-2 is most precise in the reddest pass band, achieving better than 1 ppm precision with the 650LP filter (based on four stars). At bluer wavelengths the precision is still very good -- 2.5 ppm in \sdssg{}, 6.7 ppm in 500SP. However, in the bluest band, 425SP, the precision worsens to 13.7 ppm. When used without a filter (Clear) 3.5 ppm precision is being achieved.

Observations of Sirius (HD~48915) have a systematically worse precision during the 2018MAR run than the other stars shown in table \ref{tab:prec_aat}. The reasons for this are unclear. If we remove the Sirius observations, the mean $e_p$ values for the bands are 11.6, 6.2, 1.7 and 2.9 ppm for 425SP, 500SP, \sdssg{} and Clear bands respectively.

\begin{table*}
\caption{Precision from repeat observations of bright stars with HIPPI-2 at the AAT.}
\centering
\begin{tabular}{lrc|rrrr|rrrr|rr}
\hline
\hline
Star & n & $\lambda_{eff}$ & $q\pm\Delta q$ & $\sigma_q$ & $\delta_q$ & $e_q$ & $u\pm\Delta u$ & $\sigma_u$ & $\delta_u$ & $e_u$ & $\sigma_p$ & $e_p$ \\
    &   &   (nm)        &   \multicolumn{1}{r}{(ppm)\0} &&&  & \multicolumn{1}{r}{(ppm)\0} &&&& & \\
\hline
\hline
\textit{425SP (B)} &&&&&&&&&&&&\\
HD  48915 & 3 & 401.4 & 14.6$\pm$2.2 & 34.7 & 3.9 & 34.5 & 7.9$\pm$2.2 & 14.3 & 3.9 & 13.7 & 24.5 & 24.1 \\
HD  50241 & 3 & 403.2 & 99.2$\pm$6.5 & 4.3 & 11.4 & 0.0 & 47.7$\pm$6.3 & 3.4 & 11.0 & 0.0 & 3.8 & 0.0 \\
HD  80007*& 3 & 398.4 & -45.6$\pm$2.1 & 3.7 & 3.5 & 0.0 & 14.7$\pm$2.2 & 3.8 & 13.3 & 12.8 & 8.4 & 6.4 \\
HD  97603 & 5 & 402.9 & 36.0$\pm$3.7 & 31.4 & 8.5 & 30.2 & -24.2$\pm$3.7 & 27.3 & 8.4 & 26.0 & 29.3 & 28.1 \\
HD 102647 & 3 & 402.3 & 13.0$\pm$3.8 & 7.7 & 6.5 & 4.1 & -3.4$\pm$3.7 & 21.9 & 6.4 & 20.9 & 14.8 & 12.5 \\
HD 102870 & 3 & 405.3 & -26.3$\pm$7.7 & 10.2 & 13.4 & 0.0 & -3.9$\pm$7.6 & 25.6 & 13.2 & 21.9 & 17.9 & 11.0 \\
& & 403.0 & & & & & & & & & 16.5 & 13.7 \\
\hline
\textit{500SP (B)} &&&&&&&&&&&&\\
HD  48915 & 3 & 437.7 & -6.8$\pm$0.9 & 6.7 & 1.5 & 6.6 & -2.5$\pm$0.9 & 9.4 & 1.6 & 9.2 & 8.1 & 7.9 \\
HD  97603 & 3 & 440.8 & 21.5$\pm$2.1 & 4.7 & 3.6 & 3.0 & -7.3$\pm$2.0 & 10.6 & 3.5 & 10.1 & 7.7 & 6.5 \\
HD 102647 & 3 & 439.2 & 10.2$\pm$1.6 & 8.1 & 2.8 & 7.6 & -3.4$\pm$1.7 & 7.1 & 3.0 & 6.4 & 7.6 & 7.0 \\
HD 102870 & 3 & 445.7 & -4.2$\pm$3.2 & 11.8 & 5.5 & 10.5 & 5.6$\pm$3.2 & 5.3 & 5.5 & 0.0 & 8.6 & 5.2 \\
& & 440.9 & & & & & & & & & 8.0 & 6.7 \\
\hline
\textit{\sdssg{} (B)} &&&&&&&&&&&&\\
HD  48915 & 3 & 463.4 & -10.9$\pm$0.7 & 9.2 & 1.1 & 9.2 & -3.1$\pm$0.7 & 5.6 & 1.1 & 5.4 & 7.4 & 7.3 \\
HD  48915*& 3 & 462.5 & 3.8$\pm$0.8 & 1.8 & 0.8 & 0.0 & 3.5$\pm$0.7 & 1.3 & 3.5 & 3.3 & 2.1 & 1.6 \\
HD  50241 & 5 & 466.2 & 37.6$\pm$1.2 & 2.3 & 2.7 & 0.0 & 21.9$\pm$1.2 & 4.6 & 2.7 & 3.7 & 3.4 & 1.8 \\
HD  80007 & 4 & 464.0 & -6.6$\pm$0.9 & 3.4 & 1.8 & 2.8 & 16.8$\pm$0.9 & 3.6 & 1.9 & 3.1 & 3.5 & 2.9 \\
HD  97603 & 4 & 466.2 & 20.4$\pm$1.4 & 3.9 & 2.8 & 2.7 & -9.6$\pm$1.4 & 2.5 & 2.8 & 0.0 & 3.2 & 1.3 \\
HD 102647 & 3 & 465.0 & 6.0$\pm$1.3 & 3.8 & 2.3 & 3.0 & -3.6$\pm$1.3 & 2.8 & 2.2 & 1.7 & 3.3 & 2.3 \\
HD 102870 & 3 & 470.5 & 4.8$\pm$2.6 & 2.5 & 4.4 & 0.0 & 6.7$\pm$2.6 & 0.9 & 4.5 & 0.0 & 1.7 & 0.0 \\
& & 465.6 & & & & & & & & & 3.5 & 2.5 \\
\hline
\textit{V (B)} &&&&&&&&&&&&\\
HD  48915 & 3 & 533.2 & -0.1$\pm$0.9 & 1.9 & 1.6 & 1.0 & -0.6$\pm$0.9 & 6.8 & 1.5 & 6.6 & 4.3 & 3.8 \\
\hline
\textit{\sdssr{} (R)} &&&&&&&&&&&&\\
HD 102647 & 3 & 622.6 & 2.1$\pm$2.0 & 2.6 & 3.5 & 0.0 & 3.8$\pm$2.1 & 2.1 & 3.6 & 0.0 & 2.4 & 0.0 \\
\hline
\textit{650LP (R)} &&&&&&&&&&&&\\
HD  50241 & 3 & 722.5 & -23.1$\pm$3.9 & 6.5 & 6.7 & 0.0 & 18.3$\pm$4.0 & 6.7 & 6.9 & 0.0 & 6.6 & 0.0 \\
HD  97603 & 3 & 721.9 & 8.0$\pm$3.6 & 2.5 & 6.2 & 0.0 & -4.9$\pm$3.6 & 2.5 & 6.2 & 0.0 & 2.5 & 0.0 \\
HD 102647 & 3 & 721.4 & 4.0$\pm$2.9 & 7.0 & 5.1 & 4.8 & 4.7$\pm$2.9 & 0.8 & 5.1 & 0.0 & 3.9 & 2.4 \\
HD 175191 & 3 & 719.0 & -43.7$\pm$2.8 & 2.3 & 4.8 & 0.0 & -209.5$\pm$2.7 & 5.0 & 4.8 & 1.4 & 3.6 & 0.7 \\
& & 721.2 & & & & & & & & & 4.2 & 0.8 \\
\hline
\textit{Clear (B)} &&&&&&&&&&&&\\
HD  48915 & 4 & 469.4 & -13.2$\pm$0.5 & 8.0 & 1.0 & 7.9 & -7.1$\pm$0.5 & 4.2 & 1.0 & 4.1 & 6.1 & 6.0 \\
HD  48915*& 3 & 464.3 & -1.9$\pm$1.0 & 2.0 & 4.9 & 4.5 & -0.7$\pm$1.0 & 2.0 & 2.3 & 1.0 & 3.6 & 2.8 \\
HD 102647 & 4 & 473.0 & 6.9$\pm$1.0 & 2.3 & 2.1 & 1.0 & -1.6$\pm$1.0 & 4.1 & 2.1 & 3.5 & 3.2 & 2.3 \\
HD 102870 & 3 & 489.3 & 3.8$\pm$2.4 & 4.8 & 4.1 & 2.5 & 7.8$\pm$2.3 & 0.8 & 4.0 & 0.0 & 2.8 & 1.3 \\
HD 140573 & 5 & 504.6 & 1.5$\pm$1.4 & 5.2 & 3.0 & 4.2 & 2.6$\pm$1.3 & 6.6 & 3.0 & 5.9 & 5.9 & 5.0 \\
& & 481.3 & & & & & & & & & 4.3 & 3.5 \\
\hline
\hline
\end{tabular}
\begin{flushleft}
Notes: \\
All values of $\sigma$, $\delta$ and $e$ are in ppm. \\
B and R designations given parenthetically indicate which PMT was used.\\
* Indicates 2019MAR run (ML-E1 modulator), all other observations were from the 2018JUL/AUG run (BNS-E4-7 modulator).\\
\end{flushleft}
\label{tab:prec_aat}
\end{table*}

The reported precision of HIPPI at the AAT \citep{bailey15} was 4.3 ppm based on combined measurements of $\sigma$ made in the \sdssg{} and 500SP bands; HIPPI-2 appears to be doing slightly better than this. In table \ref{tab:comparison} we compare the HIPPI-2 AAT precision measurements with a similar analysis of HIPPI data from 2014 to 2017. We have made relatively few sets of repeat observations in redder bands, so these are combined in the table to give a meaningful comparison.
It can be seen from table \ref{tab:comparison} that HIPPI-2 outperforms HIPPI for most bands in terms of both the $\sigma_p$ and $e_p$ measurements.

In part that may be due to the use of a larger aperture. Table \ref{tab:prec_ap} shows precision determinations made without a filter for the same target ($\beta$ Leo) with different aperture sizes. The precision is seen to improve with increasing aperture size. Although typically around 2\arcsec{} or better, the seeing at the AAT can often reach 5\arcsec{} and is occasionally much worse. Under such conditions a significant fraction of the light would fall outside of HIPPI's 6.9\arcsec{} aperture. Thus a larger aperture improves things for bright stars where the increased sky background is not significant. 


\begin{table}
\caption{A comparison of the precision of HIPPI and HIPPI-2 on the AAT by band.}
\centering
\begin{tabular}{l|rrr|rrr}
\hline
        &   \multicolumn{3}{c}{HIPPI-2}     & \multicolumn{3}{c}{HIPPI}         \\
Band    &   N   & $\sigma_p$    & $e_p$     &   N   & $\sigma_p$    & $e_p$     \\
\hline
425SP$^a$  &   5   &   16.5        &   13.7    &   4   &   21.2        &   13.2    \\
500SP$^a$  &   4   &  \08.0        &  \06.7    &   2   &  \09.6        &  \07.5    \\
\sdssg{}$^{ab}$& 7 &  \03.5        &  \02.5    &   3   &  \04.4        &  \02.1    \\
Clear$^a$ &   6   &  \04.3        &  \03.5    &   6   &  \06.1        &  \04.7    \\
Redder$^c$ &  6   &  \03.5        &  \01.1    &   3   &  \03.1        &  \01.6    \\
\hline
\end{tabular}
\begin{flushleft}
Notes: \\
All values of $\sigma$ and $e$ are in ppm. \\
$^a$ If we remove the Sirius observations from the HIPPI-2 results, the mean $e_p$ (N) values for the bands are 11.6 ppm (4), 6.2 ppm (3), 1.7 ppm (5) and 2.9 ppm (3) for 425SP, 500SP, \sdssg{} and Clear bands respectively. \\
$^b$ Includes observations made in two different versions of the \sdssg{} filter with HIPPI. \\
$^c$ Combined V, \sdssr{} (with both B and R PMTs) and 650LP data. \\
\end{flushleft}
\label{tab:comparison}
\end{table}

\begin{table*}
\caption{Precision in different sized aperture observations of HD 102647 made with no filter.}
\centering
\begin{tabular}{lrc|rrrr|rrrr|rr}
\hline
Aperture\0 & n & $\lambda_{\rm eff}$ & $q\pm\Delta q$	& $\sigma_q$ & $\delta_q$ & $e_q$ & $u\pm\Delta u$ & $\sigma_u$ & $\delta_u$ & $e_u$ & $\sigma_p$ & $e_p$ \\
    &   &   (nm)        &   \multicolumn{1}{r}{(ppm)\0} &&&  & \multicolumn{1}{r}{(ppm)\0} &&&& & \\
\hline
5.3\arcsec{} & 3 & 472.9 & \0-0.4$\pm$1.4 & 7.0 & 2.5 & 6.6 & 0.5$\pm$1.5 & 6.2 & 2.6 & 5.7 & 6.6 & 6.1 \\
8.6\arcsec{} & 3 & 472.9 & 4.5$\pm$1.2 & 8.0 & 2.1 & 7.7 & 1.3$\pm$1.2 & 2.9 & 2.1 & 2.0 & 5.4 & 4.8 \\
15.7\arcsec{} & 4 & 473.0 & 6.9$\pm$1.0 & \02.3 & \02.1 & \01.0 & \0\0-1.6$\pm$1.0 & \04.1 & \02.1 & \03.5 & \03.2 & \02.3 \\
\hline
\end{tabular}
\begin{flushleft}
Notes:\\
All values of $\sigma$, $\delta$ and $e$ are in ppm.
\end{flushleft}
\label{tab:prec_ap}
\end{table*}

Table \ref{tab:prec_wsu} presents precision measurements made during runs at WSU -- the 2018MAY and 2019FEB runs in Clear and with an \sdssg{} filter. It can be seen that the precision measured as either $\sigma_p$ or $e_p$ is not as good as that at the AAT.

\begin{table*}
\caption{Precision from repeat observations of bright stars with HIPPI-2 at WSU.}
\centering
\begin{tabular}{lrc|rrrr|rrrr|rr}
\hline
\hline
Star & n & $\lambda_{\rm eff}$ & $q\pm\Delta q$ & $\sigma_q$ & $\delta_q$ & $e_q$ & $u\pm\Delta u$ & $\sigma_u$ & $\delta_u$ & $e_u$ & $\sigma_p$ & $e_p$ \\
    &   &   (nm)        &   \multicolumn{1}{r}{(ppm)\0} &&&  & \multicolumn{1}{r}{(ppm)\0} &&&& & \\
\hline
\hline
\textit{\sdssg} &&&&&&&&&&&&\\
HD  48915 & 4 & 464.9 & -0.8$\pm$2.9 & 7.9 & 5.9 & 5.1 & -0.9$\pm$2.8 & 9.8 & 5.7 & 8.0 & 8.8 & 6.5 \\
HD  48915*& 4 & 463.1 & -2.8$\pm$1.7 & 9.8 & 3.7 & 9.1 & -3.2$\pm$1.6 & 13.9 & 3.6 & 13.4 & 11.9 & 11.3 \\ 
HD  80007 & 3 & 464.4 & -8.1$\pm$5.3 & 12.7 & 9.6 & 8.4 & 7.6$\pm$5.0 & 10.4 & 9.1 & 5.0 & 11.6 & 6.7 \\
& & 464.6 & & & & & & & & & 10.8 & 8.2 \\
\hline
\textit{Clear} &&&&&&&&&&&&\\
HD  48915 & 6 & 471.7 & -1.7$\pm$1.5 & 10.5 & 3.8 & 9.7 & 11.0$\pm$1.5 & 14.5 & 3.8 & 14.0 & 12.5 & 11.9 \\
HD  48915*& 3 & 467.3 & 2.1$\pm$3.3 & 10.0 & 6.2 & 7.9 & 0.6$\pm$3.2 & 3.3 & 6.0 & 0.0 & 6.7 & 3.9 \\ 
HD  80007 & 6 & 470.8 & -5.0$\pm$2.9 & 4.1 & 7.3 & 0.0 & 21.6$\pm$2.9 & 11.2 & 7.2 & 8.6 & 7.7 & 4.3 \\
HD 102647 & 5 & 475.3 & 4.0$\pm$4.1 & 9.9 & 9.2 & 3.6 & -14.5$\pm$4.1 & 17.2 & 9.1 & 14.6 & 13.6 & 9.1 \\
& & 472.6 & & & & & & & & & 10.1 & 7.3 \\
\hline
\hline
\end{tabular}
\begin{flushleft}
Notes: \\
All values of $\sigma$, $\delta$ and $e$ are in ppm.
* Indicates 2019FEB run (ML-E1 modulator), all other observations were from the 2018MAY run (BNS-E3 modulator).
\end{flushleft}
\label{tab:prec_wsu}
\end{table*}

\subsection{What Limits the Precision?}
\label{sec:limits}

Based on the results given above we can consider what is limiting the precision achievable with these FLC based instruments. We believe the main limitations are set by the instrumental polarization that is inherent in this instrument design. As discussed in \citet{bailey15} these instruments have a large (1000s of ppm) instrumental polarization which is intrinsic to the modulators. We largely eliminate this instrumental polarization by rotating the modulator relative to the rest of the instrument so that the instrumental polarization is orthogonal to the Stokes parameter being measured (This is an adjustment done when the instrument is first set up for each run). Residual effects are cancelled by the second-stage chopping procedure of repeating observations at 90 degree separated angles (see section \ref{sec:obs}). 

We suspect that there is a small spatial variation of this instrumental polarization across the modulator probably associated with the fringing effects described by \citet{gisler03}. This means that the measured polarization could be different if the star is not precisely centered in the instrument aperture. Such an effect can explain the differences between precision on different telescopes. The AAT has very good tracking and we normally autoguide using an off-axis guide star. At the WSU telescope we are not able to autoguide. On the UNSW 35 cm Celestron telescope used with Mini-HIPPI centering of objects is difficult due to backlash in the telescope drives. The precision we obtain on Mini-HIPPI measured using obervations of HD~49815 or HD~128620 in the Clear band is $\sigma_p$ = 19.8 ppm and $e_p$ = 14.0 ppm. It therefore seems likely that the poorer precision obtained with the smaller telescopes is due to poorer tracking leading to errors in centering of the stars.

\begin{figure}
\includegraphics[width=\columnwidth]{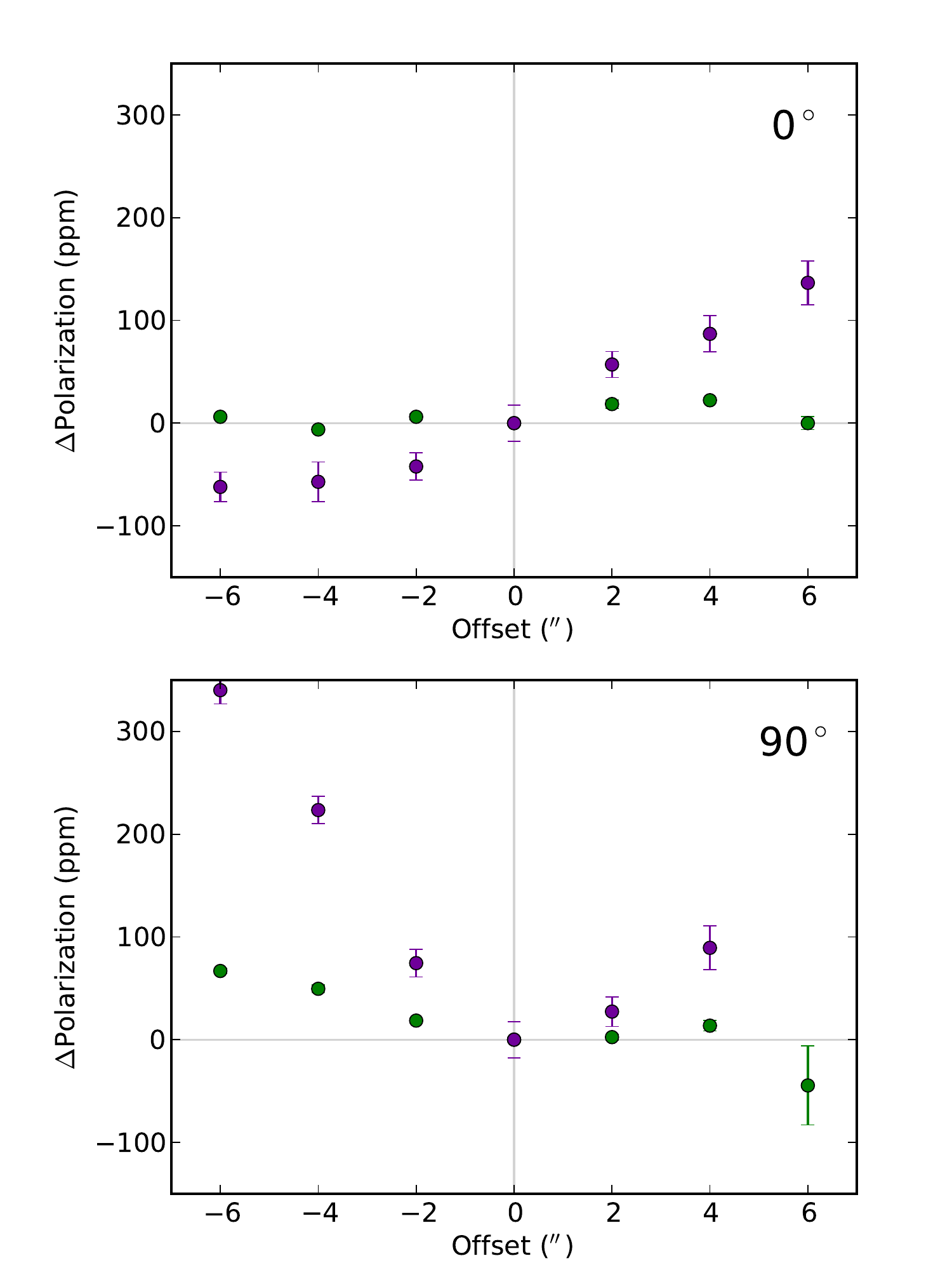}
\caption{The difference between polarization recorded from a measurement in a centered and offset position at a PA of 0$^{\circ}$ (top) and 90$^{\circ}$ (bottom). The data points are colour coded according to the filter: \sdssg{}  (green), 425SP (violet). Not shown is the 425SP value for 90$^{\circ}$ at an offset of 6\arcsec{} which was -894 $\pm$ 162 ppm, indicating it was very near the edge of the aperture.}
 \label{fig:center}
\end{figure}

During the 2018MAR AAT observing run we made a number of short measurements of Sirius to determine the effect miscentering has. For this purpose the 2.6 mm (11.9\arcsec) aperture was used, and the results read from the on-screen quick-look polarization determination. Measurements were acquired at two orthogonal PAs with both the \sdssg{}  and 425SP filter. The star was first centred in the normal way, and then off-set in 2\arcsec{} increments either side of centre. A representative efficiency correction was made to the measurements, and the results are shown in figure \ref{fig:center} as the difference between the measurement at centre and each subsequent measurement.

There is a trend such that the further off center the target is, the greater is the likely deviation from the centered value. The effect is much more pronounced in the 425SP band than the \sdssg{}  band, so this confirms our suspicions, and also helps to explain why the precision of HIPPI and HIPPI-2 is poorer at blue wavelengths.

\subsection{Performance versus magnitude}

\begin{figure}
\includegraphics[width=\columnwidth]{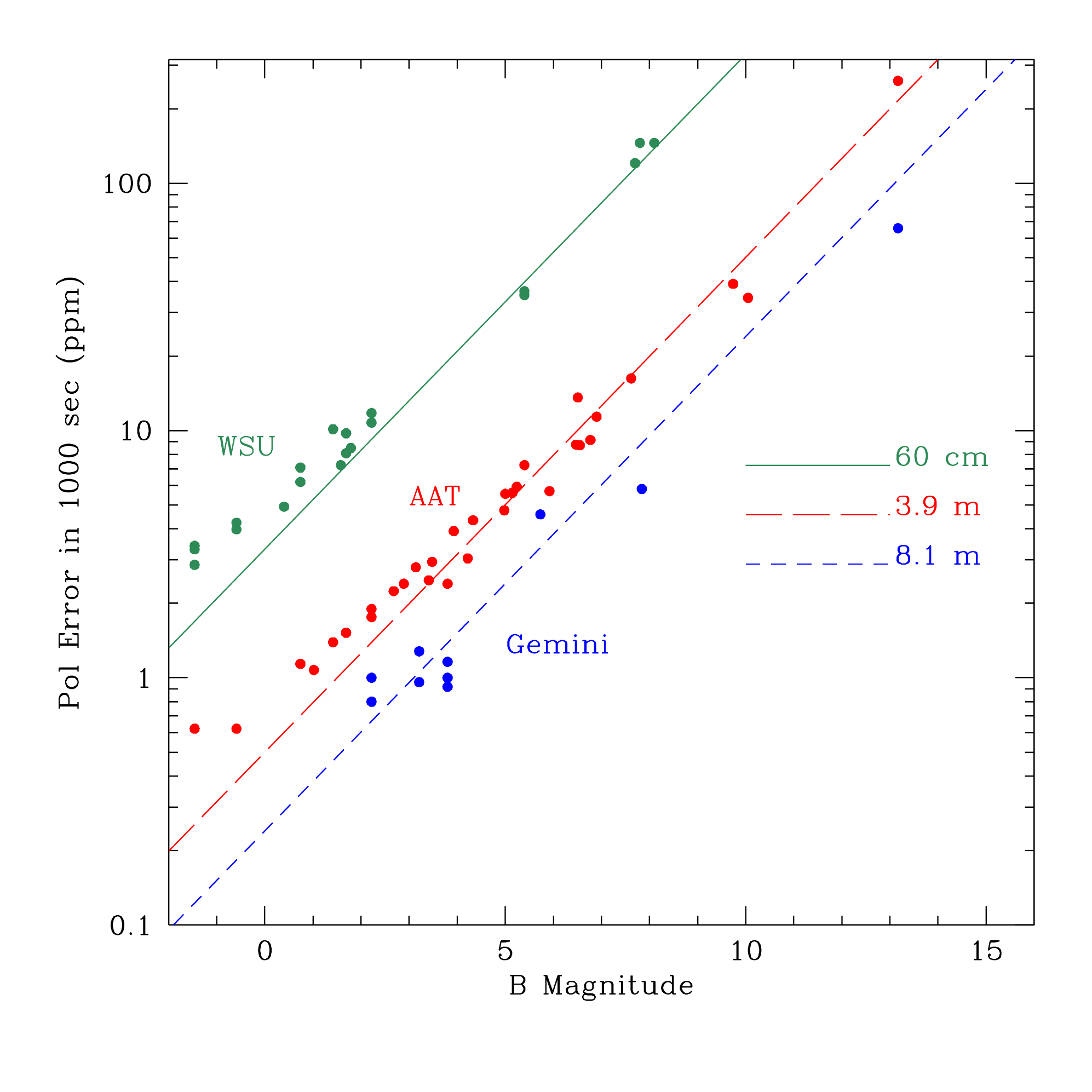}
\caption{Internal errors of observations with HIPPI-2 scaled to an integration time of 1000 seconds and plotted against B magnitude. The lines which are fitted through the fainter observations have the slope expected for photon noise limited observations and are spaced by the scaling factors expected for the change in collecting area of the three different telescopes.}
 \label{fig5}
\end{figure}

In figure \ref{fig5} we show the error in polarization, as determined by the data reduction system, plotted against the B magnitude of the object observed. For this purpose we selected observations obtained in good sky conditions in the Clear and \sdssg{}  filters. The errors have been scaled to a fixed integration time ($T$) of 1000 seconds under the assumption that the error varies as $T^{-0.5}$. The results scale with magnitude in the way expected for photon-shot-noise-limited performance as shown by the lines in the diagram. The dependence on telescope aperture is also as expected.

However, it can be seen that for very bright stars the performance is relatively poorer, with the points lying above the line. This was also seen in the similar plot for Mini-HIPPI \citep{bailey17}. Comparison of the curves for different telescopes suggest that the points start to deviate from the line at a similar magnitude (B $\sim$ 3), rather than at a similar signal level. This suggests that the effect is not due to instrumental noise sources such as those from the PMT. However, it is consistent with the idea that scintillation noise on bright stars becomes significant and is not completely removed by our 500 Hz modulation frequency \citep{bailey17}.

\subsection{Position Angle Precision}

As described in section \ref{sec:pa_corr} PA is calibrated by comparing \sdssg{} and Clear measurements of the polarized standards listed in table \ref{tab:hp_std} with their literature values to determine $\Delta PA$. The uncertainties in the literature values are typically of order a degree. Within this limitation, HIPPI-2's precision in PA can be gauged by looking at the standard deviation of $\Delta PA$ for each run; this is done in table \ref{tab:pa_prec}. 

With the exception of the 2018JUL observing run all the standard deviations fall within a degree, which is about as good as can be expected. However, HIPPI performed a little better by the same measure \citep{bailey19}. It is noteworthy that the standard deviations are largest for 2018JUL and 2018AUG when the modulator performance was drifting, and also for 2018JUN on Gemini North where the TP is very large and difficult to model (see section \ref{sec:gemini}). Without these difficulties it is reasonable to expect that PA calibration will be able to be performed as well with HIPPI-2 as with HIPPI, and that the in-run repeatability will be limited only by the precision of the rotator and rotator control software.

The determined PA for HD 203532 in \sdssg{} of 125.2 $\pm$ 0.9$^{\circ}$ from the 2018JUL run is unusually low compared to its literature value of 127.8$^{\circ}$ \citep{bagnulo17}. It was also observed in other filters during two different acquisitions, and the PA is consistent with the \sdssg{} observation. So, the observation is not a rogue, but represents a clear difference to the literature.

\begin{table}
\caption{Precision in PA by Observing Run}
\centering
\tabcolsep 1.5 pt
\begin{tabular}{lc|cccccccccccc|r}
\hline
Run & S/R &  \multicolumn{12}{c|}{Standard Observations} & S.D.\\
        &   & A & B & C & D & E & F & G & H & I & J & K & L & \multicolumn{1}{|c}{($^{\circ}$)} \\
\hline
2018JAN &   & 0 & 0 & \textit{1} & 0 & 0 & 0 & 0 & 0 & 0 & 0 & 0 & 0 & - \\
2018FEB & A & 0 & 0 & 1 & 0 & 0 & 0 & 0 & 0 & 0 & 0 & 0 & 0 & - \\
2018FEB & B & 0 & 1 & 0 & 0 & 0 & 0 & 0 & 0 & 0 & 0 & 0 & 0 & - \\
2018FEB & C & 0 & \textit{2} & 0 & 0 & 0 & 0 & 0 & 0 & 0 & 0 & 0 & 0 & 0.08 \\
2018FEB & D & 0 & 1 & 0 & 0 & 0 & 0 & 0 & 0 & 0 & 0 & 0 & 0 & - \\
2018MAR &   & 0 & 1 & 0 & 1 & 1 & 0 & 0 & 0 & 0 & 1 & 0 & 0 & 0.26 \\
2018MAY &   & 0 & 0 & 0 & 0 & \textit{2} & 0 & 0 & 0 & 0 & 0 & 0 & 0 & 0.15 \\
2018JUN &   & 0 & 0 & 0 & 0 & 1 & 0 & 1 & 0 & 1 & 0 & 0 & \textit{2} & 0.80 \\
2018JUL &   & 0 & 0 & 0 & 0 & 1 & 0 & 1 & 1 & 0 & 0 & 1 & 0 & $^a$1.56 \\
2018AUG &   & 0 & 0 & 0 & 0 & 3 & 0 & 0 & 3 & 0 & \textit{5} & 0 & 0 & 0.86 \\
2019FEB &   & 0 & 0 & 1 & 0 & 1 & 0 & 0 & 0 & 0 & 0 & 0 & 0 & 0.11 \\
2019MAR &   & 0 & 1 & 1 & 0 & 1 & 0 & 0 & 0 & 0 & 0 & 0 & 0 & 0.46 \\
\hline
\end{tabular}
\begin{flushleft}
Notes: \\
The key for the letters denoting the low polarization standards is in table \ref{tab:hp_std}. \\
All standards were observed in \sdssg{}  except the following which were observed in Clear: All from 2018JAN, 1$\times$ HD~80558 (Standard B) from 2018FEB~C, All from 2018MAY, 1$\times$ HD~120121 (Standard L) from 2018JUN, 1$\times$ HD~187929 (Standard J) from 2018AUG -- which have all been italicised in the table. \\ 
$^a$ If HD 203532 (Standard K) is excluded: 0.93. \\
\end{flushleft}
\label{tab:pa_prec}
\end{table}

\subsection{Accuracy}
\label{sec:accuracy}

The accuracy of HIPPI-2 on high polarization objects can be gauged by comparing observations made of high polarization standards with predictions made by the bandpass model. In table \ref{tab:acc} all the high polarization standard observations made with HIPPI-2 regardless of the instrument configuration of telescope are grouped by band -- the filter and PMT combination -- and the average and standard deviation of the ratio of observation to prediction reported. In this case the predictions are calculated assuming purely interstellar polarization based on the literature values given in table \ref{tab:hp_std}. It should be noted that there are reports of polarization variability in a number of these polarization standards \citep{bastien88}.

In the majority of bands the mean observed polarization is within 1.5\% of that calculated by the bandpass model. This is not surprising given that the same standards were used to calibrate the modulator curves. However, the discrepancy is larger for bands corresponding to the edge of the PMT response curves at the red end, or the rapid drop-off in modulator efficiency at the blue end. In particular it is noteworthy that the combination of the \sdssr{} filter and the blue sensitive PMT corresponds to both a comparatively large mean discrepancy, 3.3\%, and a large standard deviation of 8.3\%, while the combination of the \sdssr{} filter and the red sensitive PMT results in much more favourable measures. The most likely explanation is that the PMT response curves are not accurate at the extremes of their range. The optical components of the instrument have been characterised in the lab, but for the PMTs we rely on the manufacturer's data which makes no allowance for variance between units. Our practical experience with different PMT units leads us to believe such differences are significant.

The standard deviations given in table \ref{tab:acc} are typically around 6\% for the middle bands. It is reasonable to expect that this figure is influenced by the modulator drift, which was greatest during 2018AUG and the large TP on Gemini North during 2018JUN. We also noted that HD~149757 ($\zeta$ Oph -- an Oe star) displayed short term variability, and that the observation to prediction ratio for HD~203532 was typically high by 10\% or more in each band. Thus an analysis of the \sdssg{} band with these observations removed was also carried out. The result is a drop in the standard deviation by 1.5\% to 4.5\%. This figure is better, but is still limited by the accuracy of the literature polarization data for the standards. All of these will have been acquired with less precise instrumentation. 

\begin{table*}
\caption{Accuracy by Band}
\centering
\begin{tabular}{lcl|cccccccccccc|rr}
\hline
\hline
\multicolumn{2}{c}{Band} & $\lambda_{\rm eff}$  &  \multicolumn{12}{c|}{Standard Observations}      & \multicolumn{2}{|c}{Obs./Pred.}\\
Fil & PMT & (nm)  & A & B & C & D & E & F & G & H & I & J & K & L & Mean & S.D. \\
\hline
\hline
\multicolumn{3}{l|}{\textit{All Observations}} &&&&&&&&&&&&&&\\ 
U & B     & 381.7 & 0 & 1 & 0 & 0 & 1 & 0 & 0 & 0 & 0 & 0 & 0 & 0 & 1.124 & 0.004 \\
425SP & B & 404.2 & 0 & 1 & 0 & 0 & 3 & 1 & 1 & 2 & 1 & 3 & 1 & 1 & 1.031 & 0.081 \\
500SP & B & 441.5 & 0 & 0 & 0 & 0 & 2 & 0 & 2 & 2 & 1 & 2 & 1 & 1 & 1.013 & 0.058 \\
\sdssg{} & ${}^B{\mskip -5mu/\mskip -3mu}_R$
          & 473.9 & 0 & 5 & 3 & 1 & 8 & 1 & 2 & 4 & 1 & 6 & 1 & 1 & 0.995 & 0.060 \\
Clear & B & 481.6 & 0 & 1 & 1 & 0 & 2 & 3 & 0 & 0 & 0 & 1 & 0 & 1 & 1.015 & 0.063 \\
V &  ${}^B{\mskip -5mu/\mskip -3mu}_R$ 
          & 541.0 & 0 & 0 & 0 & 0 & 1 & 1 & 1 & 2 & 0 & 2 & 1 & 0 & 1.001 & 0.066 \\
500SP & R & 552.2 & 0 & 0 & 0 & 0 & 1 & 0 & 0 & 0 & 0 & 1 & 0 & 0 & 0.988 & 0.031 \\
\sdssr{}&B& 605.0 & 0 & 0 & 0 & 0 & 3 & 0 & 3 & 2 & 1 & 2 & 1 & 1 & 1.033 & 0.083 \\
\sdssr{}&R& 626.3 & 0 & 1 & 0 & 0 & 1 & 1 & 0 & 1 & 0 & 1 & 0 & 0 & 0.997 & 0.023 \\
425SP & R & 714.5 & 0 & 0 & 0 & 0 & 1 & 0 & 0 & 1 & 0 & 1 & 0 & 0 & 0.923 & 0.021 \\
650LP & R & 730.2 & 0 & 1 & 0 & 0 & 1 & 1 & 0 & 1 & 0 & 1 & 0 & 0 & 1.049 & 0.031 \\
\hline
\multicolumn{3}{l|}{\textit{Selected Observations}} &&&&&&&&&&&&&&\\
\sdssg{} & ${}^B{\mskip -5mu/\mskip -3mu}_R$
          & 472.7 & 0 & 5 & 3 & 1 & 4 & 0 & 1 & 1 & 0 & 2 & 0 & 0 & 1.011 & 0.045 \\
\hline
\hline
\end{tabular}
\begin{flushleft}
Notes: \\
The Mean and standard deviation (S.D.) are in ratio units.
The key for the letters denoting the low polarization standards is in table \ref{tab:hp_std}. \\
Selected observations exclude runs 2018JUN and 2018AUG, HD~149757 and HD~203532.\\
\end{flushleft}
\label{tab:acc}
\end{table*}

\subsection{Gemini North Observations}
\label{sec:gemini}

Some adjustments need to be made to the correction procedure when observing on a telescope with an AltAz mount. This is because the orientation of the telescope tube and mirrors, and hence the telescope polarization, relative to the sky systematically varies with parallactic angle, $\theta$. In the ideal case, $q$ and $u$ for any given observation will be given by \begin{equation}q=p_{TP}\cos (2\theta-\theta_{TP}) +q_{\star}+p_i,\label{eq:tpq}\end{equation} \begin{equation}u=p_{TP}\sin (2\theta-\theta_{TP}) +u_{\star}+p_i,\label{eq:tpu}\end{equation} where $p_{TP}$ is the magnitude of the telescope polarization, $\theta_{tp}$ is the parallactic angle that maximises $q_{TP}$, and $q_{\star}$ and $u_{\star}$ refer to the TP subtracted polarization of the target in the instrument frame. The instrumental component $p_i$ is largely eliminated by measuring each Stokes parameter at opposite PAs\footnote{It should be noted that a slight misalignment of the aperture with the instrument rotator will, since we are not re-centering at each PA, result in a slightly different area of the modulator being used and therefore a difference in $p_i$ between angles. Any residual in this quantity gets incorporated into the TP on an equatorial telescope. On an AltAz mount it will manifest as noise in our data if not explicitly corrected for. However, this is likely to be very small, as since adopting an observing scheme without re-centering at each PA with HIPPI-2 on the AAT we have actually measured lower TP values, compared to those seen with HIPPI, see figure \ref{fig:AATf8TP}. Similarly, a non-linear response in $p_i$ would also result in incomplete cancellation in variable conditions; something we did not see in tests made under variable cloud \citep{cotton16}.} (e.g. 0 and 90). By making multiple observations of a star at different parllactic angles it is possible to disentangle the star and telescope polarization, as was done with PlanetPol on the WHT \citep{lucas09}.

\begin{figure}
\centering
\includegraphics[width=\columnwidth]{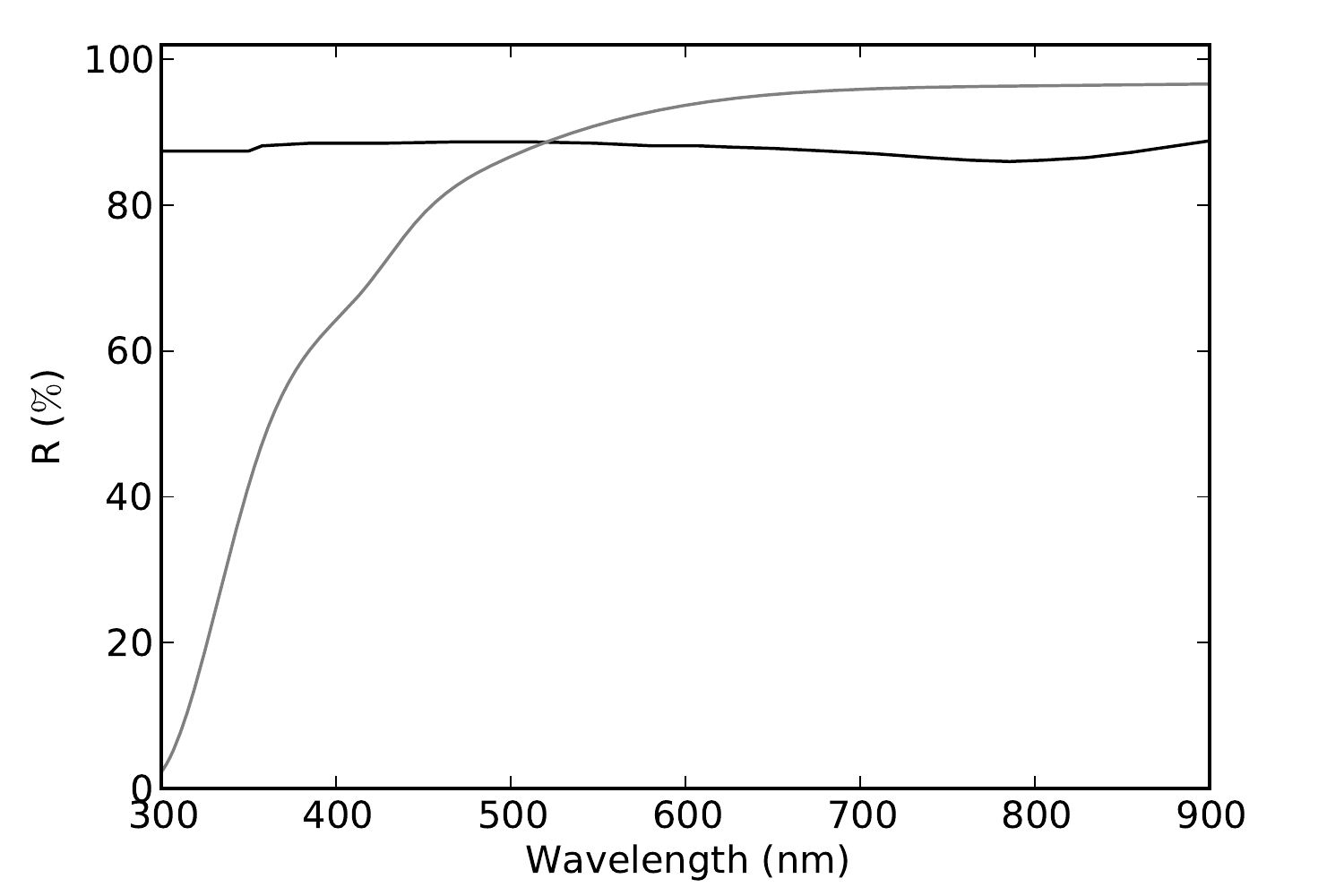}
\caption{Reflectance of the Gemini silver coated mirror (grey) compared to an aluminum coating (black).}
 \label{fig:mirrors}
\end{figure}

On Gemini North the TP was far greater than on the AAT or WSU telescopes; it was also highly wavelength dependant, being much larger at blue wavelengths. Gemini uses protected silver mirror coatings \citep{boccas04,vucina06} whereas the other telescopes we have used have standard aluminium coatings. The silver coatings provide very high reflectance at red and infrared wavelengths but the reflectance falls off in the blue and UV (see figure \ref{fig:mirrors}). The steep rise in TP we find occurs at the wavelengths where the mirror reflectance is declining.

The combination of high and strongly wavelength dependent TP prevented HIPPI-2 from obtaining the same precision it does on other telescopes. While it should be theoretically possible to subtract out all of the TP with a precise wavelength solution, the scale of the TP magnifies many issues that would otherwise be insignificant. Any imprecision in the characterisation of the optical components and detectors becomes problematic, as does the smallest misalignment of the field derotator. Any imprecision in the instrument rotator alignment would also be exacerbating. 

\begin{table}
\caption{Low polarization standard stars observed at Gemini North.}
\centering
\tabcolsep 3 pt
\begin{tabular}{lccr|cccccccc}
\hline
\hline
\multicolumn{2}{c}{Band}& \multicolumn{1}{c}{Ap} & $\lambda_{\rm eff}$ &   \multicolumn{8}{c}{Standard Observations}\\
Fil  & PMT  &\multicolumn{1}{c}{(\arcsec)}&(nm)& A & B & C & D & E & F & G & H\\
\hline
\hline
425SP    & B &  6.4 & 402.9 & 0 & 0 & 0 & 0 & 0 & 4 & 0 & 0 \\
500SP    & B &  6.4 & 443.7 & 0 & 0 & 0 & 2 & 0 & 4 & 0 & 0 \\
\sdssg{} & B &  6.4 & 475.7 & 0 & 0 & 0 & 2 & 0 & 3 & 0 & 4 \\
Clear    & B &  6.4 & 483.2 & 0 & 0 & 0 & 2 & 0 & 4 & 0 & 0 \\
\sdssr{} & B &  6.4 & 605.7 & 0 & 0 & 0 & 0 & 0 & 3 & 0 & 4 \\
\hline
\hline
\end{tabular}
\begin{flushleft}
Notes: \\
The key for the letters denoting the low polarization standards is in table \ref{tab:lp_std}. \\
\end{flushleft}
\label{tab:lp_gem}
\end{table}

\begin{figure}
\centering
\includegraphics[width=\columnwidth, trim={0cm 0.6cm 0cm 1cm},clip]{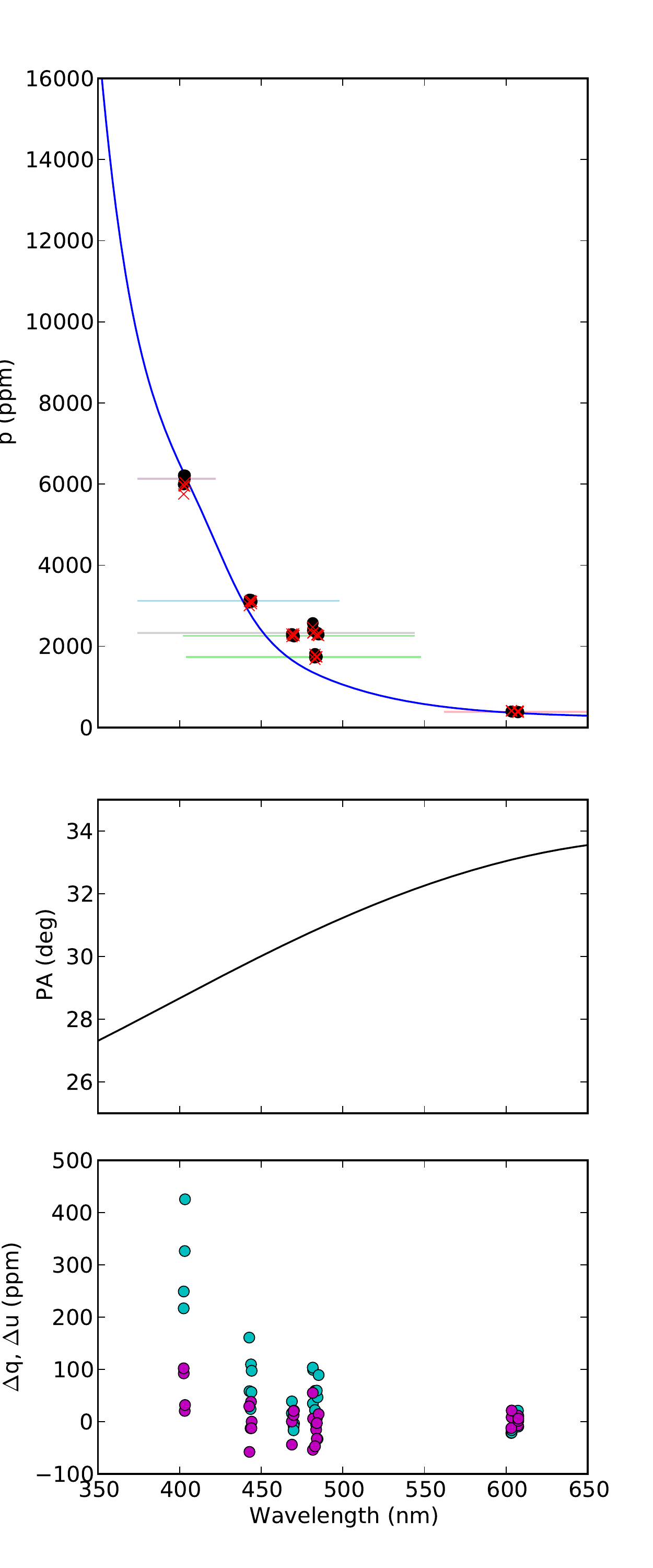}
\caption{The first order TP solution for Gemini North: in the top panel is the best fit solution in p (blue line), along with the corresponding band determinations for each low polarisation standard observation at the calculated effective wavelength (black circles), and the actual measurements (red crosses). The horizontal coloured lines are representative of the band contribution (FW10\%M); the second green line has been added for the redder standard, $\alpha$ Ser. In the middle panel is the fitted position angle of the TP (black line). The lower panel shows the residuals in $q$ (cyan circles) and $u$ (magenta circles) at the effective wavelength of the observations in the instrument frame. A second order correction is later applied to each band individually.}
 \label{fig:tp_gem}
\end{figure}

Table \ref{tab:lp_gem} indicates the low polarization standard observations made on Gemini North during the 2018JUN run. Using these observations a first order correction to the TP has been calculated by assuming that $p_{\star}$, $u_{\star}$ and $p_i$ in equations \ref{eq:tpq} and \ref{eq:tpu} are negligible, i.e. $q_{1st}=q_{TP}(\lambda)$ and $u_{1st}=u_{TP}(\lambda)$. The best fit wavelength solution is shown in figure \ref{fig:tp_gem}, along with the observations. In order to achieve the best fit in this instance a number of modifications were made to the procedure described in section \ref{sec:red&cal}. The airmass, parallactic angle and efficiency correction are calculated (at floating point precision) not just for the whole observation, but for each integration. The TP function is then calculated for each integration for each Stokes parameter in the instrumental frame, with the mean $q$ and $u$ determined for each observation. This step is especially necessary for observations made close to or through zenith. Additionally, Kurucz models selected for the temperature of the standards, rather than their spectral type were used since they provided a slightly better fit to the data. 

The PA correction is only applied after the first order TP correction is determined and subtracted, since the TP is large enough to influence the PA determined using the high polarization standards observed (see table \ref{tab:hp_std}).

The best fit function in figure \ref{fig:tp_gem} was arrived at through trial and error of many different functions and is based on assuming that the increase in TP is related to the fall-off in mirror reflectivity as discussed above. it has the form \begin{equation}p_{TP}(\lambda) = m(100-R_M(\lambda))^2 + b\end{equation} where $R_M(\lambda)$ is the mirror reflectivity as a function of wavelength in percent, and $m$ and $b$ are constants. 

The data for $R_M(\lambda)$ are taken from \citet{feller12} who present measurements of five mirror samples coated according to the Gemini recipe, a lab reference, one prior to cleaning in 2009, and three post cleaning in 2009, 2010 and 2011 which we used to create a grid. We linearly interpolated in this grid of spectra based on two parameters, the age $a$ and condition $c$ (where $c$ = 0 after cleaning and $c$ = 1 before cleaning). The values of the parameters $a=2009.0$  and $c=0.2631$ as well as $m=4.9474$ and $b=175.9$ were determined by fitting the standard star observations using Python's CURVE\_FIT routine.



In addition we also simultaneously fit a polynomial to describe the TP position angle as a function of wavelength 

\begin{equation}
PA_{tp}(\lambda)=\frac{\alpha}{\lambda}+\beta+\gamma\lambda+\delta\lambda^2+\epsilon\lambda^3
\end{equation} 

where $\lambda$ is given in nanometers and the co-efficients denoted by the greek letters are fit to be $-728.96$, $4696.4$, $-18.0912$, $0.11829934$ and $-7.68615733\times10^{-5}$ respectively.

The resultant function is shown in the middle panel of figure \ref{fig:tp_gem}.


\begin{table*}
\caption{TP and instrumental corrections based on low polarization standard observations at Gemini North.}
\centering
\tabcolsep 3 pt
\begin{tabular}{lr|rrr|rrrr|r|r}
\hline
Filter  & $\lambda_{eff}$   &  \multicolumn{3}{|c}{1st Order Corr.$^a$}   & \multicolumn{4}{|c}{2nd Order Corr.} & \multicolumn{1}{|c}{Pos Err.$^b$} & \multicolumn{1}{|c}{Prec.$^c$} \\
        & (nm)              &   \multicolumn{1}{c}{Obs. $p$}    &   \multicolumn{1}{c}{Fit $p_{TP}$}   &   \multicolumn{1}{c}{Fit $\theta_{TP}$}&   \multicolumn{1}{|c}{$q_{TP}$} & \multicolumn{1}{c}{$u_{TP}$} & \multicolumn{1}{c}{$q_{i}$} & \multicolumn{1}{c}{$u_{i}$} & \multicolumn{1}{|c|}{$e_p$} &  \\
\hline
425SP    & 402.9    & 5925.1 & 6133.3 & 28.73  & -120.4 $\pm$ 34.8 &  58.3 $\pm$ 33.9 & 261.2 $\pm$ 34.4 & -30.5 $\pm$ 34.3 & 132.2 & 140.8\\
500SP    & 443.7    & 3084.7 & 3119.7 & 29.82  &  43.9 $\pm$ 23.0 & -27.2 $\pm$ 23.6 & 114.1 $\pm$ 32.2 &  -3.2 $\pm$ 31.4 &  11.1 &  41.0\\
\sdssg{} & 475.7    & 2039.5 & 2042.9 & 30.61  &  18.5 $\pm$ 14.3 &  -8.5 $\pm$ 14.3 &  23.5 $\pm$ 19.7 &   5.1 $\pm$ 19.4 &   5.7 &  24.9\\
Clear    & 483.2    & 2328.7 & 2389.6 & 30.68  &  -1.3 $\pm$ 19.0 & -36.1 $\pm$ 18.2 &  63.8 $\pm$ 23.0 &  -9.6 $\pm$ 22.4 &  13.2 &  32.2\\
\sdssr{} & 605.7    &\0399.4 &\0386.1 & 33.06  &   8.9 $\pm$\04.8 & \04.6 $\pm$  4.8 & -10.5 $\pm$\08.3 &   7.2 $\pm$\08.2 &   1.1 &   9.7\\
\hline
\end{tabular}
\begin{flushleft}
Notes: \\
All values of $q$, $u$, $p$, $e$ and precision are in ppm. \\
The $q$ and $u$ are given in the instrument frame, and need to be rotated by 58.1$^{\circ}$ for the equatorial frame.\\
$^a$ Shown are the average measured polarizations of the standards in each band, and the average fits of $p_{TP}$ and $\theta_{TP}$ using the bandpass model. $\theta_{TP}$ represents the PA of the TP as measured for a parallactic angle of 0$^{\circ}$ as measured in the instrumental frame.\\
$^b$ The positioning error $e_p$ is calculated in the same way as in tables \ref{tab:prec_aat} and \ref{tab:prec_wsu}, i.e. having taken account of the errors in the other quantities to the left in the table. In addition to the low polarization standards, repeat observations of two other stars thought to have low levels of constant polarization were also considered. \\
$^c$ The final precision estimate considers all the stated errors as the square root of the sum of the means squared. \\
\end{flushleft}
\label{tab:gem_tp}
\end{table*}

As shown in figure \ref{fig:tp_gem}, after the subtraction of $p_{TP}(\lambda)$ there are still significant residuals in the low polarization standard measurements, where the disagreement is largest in the two shortpass bands (and appears more down to the fit $PA$ than the fit $p$). A second order correction has been applied to each band individually. This correction takes the form \begin{equation}q_{2nd}=q_{TP}+q_i\end{equation}\begin{equation}u_{2nd}=u_{TP}+u_i\end{equation} where the $q_{TP}$ and $u_{TP}$ are functions of the parallactic angle, as in equation \ref{eq:tpq} and \ref{eq:tpu}. By fitting these equations using the PYTHON package SCIPY's CURVE\_FIT routine \citep{scipy} an error in each term is obtained which allows the precision of the instrument on the telescope to be quantified. The corrections and associated errors are shown in table \ref{tab:gem_tp}. As these values dwarfed the expected values of $q_{\star}$ and $u_{\star}$ for the standards, attempts to retrieve them were abandoned. 

\section{CONCLUSIONS}

We have built and tested a new versatile and compact high-precision polarimeter HIPPI-2. The instrument is based on a Ferro-electric Liquid Crystal modulator as used in its predecessor HIPPI. HIPPI-2 is constructed largely by 3D printing. It weighs about 4~kg and requires a single compact electronics box weighing 1.3~kg containing its data acquisition electronics and computer. The new instrument has been tested on three telescopes, the 60-cm Ritchey-Chretien telescope at Western Sydney University's Penrith Observatory, the 3.9-m Anglo-Australian Telescope (AAT) and the 8.1-m Gemini North Telescope.

On the AAT HIPPI-2 achieves a precision as measured from repeat observations of low-polarization stars in the \sdssg{} band of better than 3.5 ppm and probably around 2-3 ppm. Precision is somewhat better at red wavelengths and poorer at blue wavelengths. On the WSU telescope the precision in the \sdssg{} band is better than 11 ppm and probably around 7-8 ppm. We believe the limit on the precision is set by the accuracy with which stars can be kept centered on the instrument axis, and thus the better precision with the AAT reflects its more accurate tracking and guiding.

The telescope polarization measured at the WSU telescope ranged from 10--40 ppm. The telescope polarization at the AAT ranged from 10--300 ppm with the highest values occurring with an f/15 secondary that had not been realuminised for 20 years. When this mirror was recoated much lower values (around 10 ppm in most bands) were obtained. These telescopes use standard aluminium mirror coatings.

On the Gemini North telescope, which uses protected silver mirror coatings, we found much higher, and strongly wavelength dependent telescope polarization, increasing from $\sim$400 ppm in the \sdssr{} band to $\sim$6000 ppm at 400 nm. While we have developed a model to correct for the high TP, the strong wavelength dependence introduces uncertainties that limit the precision to $\sim$10 ppm at the \sdssr{} band, $\sim$25 ppm at the \sdssg{} band, and much worse at bluer bands.

On the AAT HIPPI-2 provides improved precision, throughput and observing efficiency compared with its predecessor, HIPPI, which has already proven to be a very scientifically productive instrument. The compact size allows HIPPI-2 to be easily adapted to a range of telescopes including relatively small telescopes such as the 60-cm WSU telescope.

\begin{acknowledgements}
We thank the former Director of the Australian Astronomical Observatory, Prof. Warrick Couch, and the current Director of Siding Spring Observatory, A/Prof. Chris Lidman for their support of the HIPPI-2 project on the AAT. We thank Prof. Miroslav Filipovic for providing access to the Penrith Observatory. Funding for the construction of HIPPI-2 was provided by UNSW through the Science Faculty Research Grants Program. Nicholas Borsato, Dag Evensberget, Behrooz Karamiqucham, Fiona Lewis, Shannon Melrose and Jinglin Zhao assisted with observations at the AAT. Based on observations under program GN-2018A-DD-108, obtained at the Gemini Observatory, which is operated by the Association of Universities for Research in Astronomy, Inc., under a cooperative agreement with the NSF on behalf of the Gemini partnership: the National Science Foundation (United States), the National Research Council (Canada), CONICYT (Chile), Ministerio de Ciencia, Tecnolog\'{i}a e Innovaci\'{o}n Productiva (Argentina), and Minist\'{e}rio da Ci\^{e}ncia, Tecnologia e Inova\c{c}\~{a}o (Brazil). This research has made use of the SIMBAD database, operated at CDS, Strasbourg, France.

\end{acknowledgements}

\bibliographystyle{pasa-mnras}
\bibliography{hippi2}

\begin{thebibliography}{}
\makeatletter
\relax
\def\mn@urlcharsother{\let\do\@makeother \do\$\do\&\do\#\do\^\do\_\do\%\do\~}
\definecolor{darkblue}{rgb}{0,0,0.597656}
\def\mndoi{\begingroup\mn@urlcharsother \@ifnextchar [ {\mndoi@} {\mndoi@[]}}
\def\mndoi@[#1]#2{\def\@tempa{#1}\ifx\@tempa\@empty \href
  {http://dx.doi.org/#2} {\textcolor{darkblue}{doi:#2}}\else \href
  {http://dx.doi.org/#2} {\textcolor{darkblue}{#1}}\fi \endgroup}
\def\mn@eprint#1#2{\mn@eprint@#1:#2::\@nil}
\def\mn@eprint@arXiv#1{\href {http://arxiv.org/abs/#1} {{\tt arXiv:#1}}}
\def\mn@eprint@dblp#1{\href {http://dblp.uni-trier.de/rec/bibtex/#1.xml}
  {dblp:#1}}
\def\mn@eprint@#1:#2:#3:#4\@nil{\def\@tempa {#1}\def\@tempb {#2}\def\@tempc
  {#3}\ifx \@tempc \@empty \let \@tempc \@tempb \let \@tempb \@tempa \fi \ifx
  \@tempb \@empty \def\@tempb {arXiv}\fi \@ifundefined
  {mn@eprint@\@tempb}{\@tempb:\@tempc}{\expandafter \expandafter \csname
  mn@eprint@\@tempb\endcsname \expandafter{\@tempc}}}

\bibitem[\protect\citeauthoryear{{Astropy Collaboration} et~al.,}{{Astropy
  Collaboration} et~al.}{2013}]{astropy1}
{Astropy Collaboration} et~al., 2013, \mndoi [\aap]
  {10.1051/0004-6361/201322068}, \href
  {https://ui.adsabs.harvard.edu/abs/2013A%26A...558A..33A} {558, A33}

\bibitem[\protect\citeauthoryear{{Astropy Collaboration} et~al.,}{{Astropy
  Collaboration} et~al.}{2018}]{astropy2}
{Astropy Collaboration} et~al., 2018, \mndoi [\aj] {10.3847/1538-3881/aabc4f},
  \href {https://ui.adsabs.harvard.edu/abs/2018AJ....156..123A} {156, 123}

\bibitem[\protect\citeauthoryear{{Bagnulo} et~al.,}{{Bagnulo}
  et~al.}{2017}]{bagnulo17}
{Bagnulo} S.,  et~al., 2017, \mndoi [\aap] {10.1051/0004-6361/201731459}, \href
  {https://ui.adsabs.harvard.edu/\#abs/2017A&A...608A.146B} {608, A146}

\bibitem[\protect\citeauthoryear{{Bailey} \& {Kedziora-Chudczer}}{{Bailey} \&
  {Kedziora-Chudczer}}{2012}]{bailey12}
{Bailey} J.,  {Kedziora-Chudczer} L.,  2012, \mndoi [\mnras]
  {10.1111/j.1365-2966.2011.19845.x}, \href
  {http://adsabs.harvard.edu/abs/2012MNRAS.419.1913B} {419, 1913}

\bibitem[\protect\citeauthoryear{{Bailey}, {Lucas}  \& {Hough}}{{Bailey}
  et~al.}{2010}]{bailey10}
{Bailey} J.,  {Lucas} P.~W.,   {Hough} J.~H.,  2010, \mndoi [\mnras]
  {10.1111/j.1365-2966.2010.16634.x}, \href
  {http://adsabs.harvard.edu/abs/2010MNRAS.405.2570B} {405, 2570}

\bibitem[\protect\citeauthoryear{{Bailey}, {Kedziora-Chudczer}, {Cotton},
  {Bott}, {Hough}  \& {Lucas}}{{Bailey} et~al.}{2015}]{bailey15}
{Bailey} J.,  {Kedziora-Chudczer} L.,  {Cotton} D.~V.,  {Bott} K.,  {Hough}
  J.~H.,   {Lucas} P.~W.,  2015, \mndoi [\mnras] {10.1093/mnras/stv519}, \href
  {http://adsabs.harvard.edu/abs/2015MNRAS.449.3064B} {449, 3064}

\bibitem[\protect\citeauthoryear{{Bailey}, {Cotton}  \&
  {Kedziora-Chudczer}}{{Bailey} et~al.}{2017}]{bailey17}
{Bailey} J.,  {Cotton} D.~V.,   {Kedziora-Chudczer} L.,  2017, \mndoi [\mnras]
  {10.1093/mnras/stw2886}, \href
  {http://adsabs.harvard.edu/abs/2017MNRAS.465.1601B} {465, 1601}

\bibitem[\protect\citeauthoryear{{Bailey}, {Cotton}, {Kedziora-Chudczer}, {De
  Horta}  \& {Maybour}}{{Bailey} et~al.}{2019}]{bailey19}
{Bailey} J.,  {Cotton} D.~V.,  {Kedziora-Chudczer} L.,  {De Horta} A.,
  {Maybour} D.,  2019, \mndoi [Nature Astronomy] {10.1038/s41550-019-0738-7},
  \href {https://ui.adsabs.harvard.edu/abs/2019NatAs.tmp..240B} {3, 636}

\bibitem[\protect\citeauthoryear{{Bastien}, {Drissen}, {Menard}, {Moffat},
  {Robert}  \& {St-Louis}}{{Bastien} et~al.}{1988}]{bastien88}
{Bastien} P.,  {Drissen} L.,  {Menard} F.,  {Moffat} A.~F.~J.,  {Robert} C.,
  {St-Louis} N.,  1988, \mndoi [\aj] {10.1086/114688}, \href
  {https://ui.adsabs.harvard.edu/abs/1988AJ.....95..900B} {95, 900}

\bibitem[\protect\citeauthoryear{{Boccas}, {Vucina}, {Araya}, {Vera}  \&
  {Ahhee}}{{Boccas} et~al.}{2004}]{boccas04}
{Boccas} M.,  {Vucina} T.,  {Araya} C.,  {Vera} E.,   {Ahhee} C.,  2004, in
  {Atad-Ettedgui} E.,  {Dierickx} P.,  eds,  \procspie Vol. 5494, Optical
  Fabrication, Metrology, and Material Advancements for Telescopes. pp
  239--253, \mndoi{10.1117/12.548809}

\bibitem[\protect\citeauthoryear{{Bott}, {Bailey}, {Kedziora-Chudczer},
  {Cotton}, {Lucas}, {Marshall}  \& {Hough}}{{Bott} et~al.}{2016}]{bott16}
{Bott} K.,  {Bailey} J.,  {Kedziora-Chudczer} L.,  {Cotton} D.~V.,  {Lucas}
  P.~W.,  {Marshall} J.~P.,   {Hough} J.~H.,  2016, \mndoi [\mnras]
  {10.1093/mnrasl/slw046}, \href
  {http://adsabs.harvard.edu/abs/2016MNRAS.459L.109B} {459, L109}

\bibitem[\protect\citeauthoryear{{Bott}, {Bailey}, {Cotton},
  {Kedziora-Chudczer}, {Marshall}  \& {Meadows}}{{Bott} et~al.}{2018}]{bott18}
{Bott} K.,  {Bailey} J.,  {Cotton} D.~V.,  {Kedziora-Chudczer} L.,  {Marshall}
  J.~P.,   {Meadows} V.~S.,  2018, \mndoi [\aj] {10.3847/1538-3881/aaed20},
  \href {http://adsabs.harvard.edu/abs/2018AJ....156..293B} {156, 293}

\bibitem[\protect\citeauthoryear{{Cardelli}, {Clayton}  \& {Mathis}}{{Cardelli}
  et~al.}{1989}]{cardelli89}
{Cardelli} J.~A.,  {Clayton} G.~C.,   {Mathis} J.~S.,  1989, \mndoi [\apj]
  {10.1086/167900}, \href {http://adsabs.harvard.edu/abs/1989ApJ...345..245C}
  {345, 245}

\bibitem[\protect\citeauthoryear{{Castelli} \& {Kurucz}}{{Castelli} \&
  {Kurucz}}{2004}]{castelli&kurucz2004}
{Castelli} F.,  {Kurucz} R.~L.,  2004, ArXiv Astrophysics e-prints, \href
  {http://adsabs.harvard.edu/abs/2004astro.ph..5087C} {}

\bibitem[\protect\citeauthoryear{{Cotton}, {Bailey}, {Kedziora-Chudczer},
  {Bott}, {Lucas}, {Hough}  \& {Marshall}}{{Cotton} et~al.}{2016a}]{cotton16}
{Cotton} D.~V.,  {Bailey} J.,  {Kedziora-Chudczer} L.,  {Bott} K.,  {Lucas}
  P.~W.,  {Hough} J.~H.,   {Marshall} J.~P.,  2016a, \mndoi [\mnras]
  {10.1093/mnras/stv2185}, \href
  {http://adsabs.harvard.edu/abs/2016MNRAS.455.1607C} {455, 1607}

\bibitem[\protect\citeauthoryear{{Cotton}, {Bailey}, {Kedziora-Chudczer},
  {Bott}, {Lucas}, {Hough}  \& {Marshall}}{{Cotton} et~al.}{2016b}]{cotton16b}
{Cotton} D.~V.,  {Bailey} J.,  {Kedziora-Chudczer} L.,  {Bott} K.,  {Lucas}
  P.~W.,  {Hough} J.~H.,   {Marshall} J.~P.,  2016b, \mndoi [\mnras]
  {10.1093/mnras/stw978}, \href
  {http://adsabs.harvard.edu/abs/2016MNRAS.460...18C} {460, 18}

\bibitem[\protect\citeauthoryear{{Cotton}, {Bailey}, {Howarth}, {Bott},
  {Kedziora-Chudczer}, {Lucas}  \& {Hough}}{{Cotton} et~al.}{2017a}]{cotton17a}
{Cotton} D.~V.,  {Bailey} J.,  {Howarth} I.~D.,  {Bott} K.,
  {Kedziora-Chudczer} L.,  {Lucas} P.~W.,   {Hough} J.~H.,  2017a, \mndoi
  [Nature Astronomy] {10.1038/s41550-017-0238-6}, \href
  {http://cdsads.u-strasbg.fr/abs/2017NatAs...1..690C} {1, 690}

\bibitem[\protect\citeauthoryear{{Cotton}, {Marshall}, {Bailey},
  {Kedziora-Chudczer}, {Bott}, {Marsden}  \& {Carter}}{{Cotton}
  et~al.}{2017b}]{cotton17b}
{Cotton} D.~V.,  {Marshall} J.~P.,  {Bailey} J.,  {Kedziora-Chudczer} L.,
  {Bott} K.,  {Marsden} S.~C.,   {Carter} B.~D.,  2017b, \mndoi [\mnras]
  {10.1093/mnras/stx068}, \href
  {http://cdsads.u-strasbg.fr/abs/2017MNRAS.467..873C} {467, 873}

\bibitem[\protect\citeauthoryear{{Cotton} et~al.,}{{Cotton}
  et~al.}{2019a}]{cotton19a}
{Cotton} D.~V.,  et~al., 2019a, \mndoi [\mnras] {10.1093/mnras/sty3180}, \href
  {http://adsabs.harvard.edu/abs/2019MNRAS.483.1574C} {483, 1574}

\bibitem[\protect\citeauthoryear{{Cotton} et~al.,}{{Cotton}
  et~al.}{2019b}]{cotton19b}
{Cotton} D.~V.,  et~al., 2019b, \mndoi [\mnras] {10.1093/mnras/sty3318}, \href
  {http://adsabs.harvard.edu/abs/2019MNRAS.483.3636C} {483, 3636}

\bibitem[\protect\citeauthoryear{{Feller}, {Krishnappa}, {Pleier},
  {Hirzberger}, {Jobst}  \& {Sch{\"u}rmann}}{{Feller} et~al.}{2012}]{feller12}
{Feller} A.,  {Krishnappa} N.,  {Pleier} O.,  {Hirzberger} J.,  {Jobst} P.~J.,
   {Sch{\"u}rmann} M.,  2012, in Modern Technologies in Space- and Ground-based
  Telescopes and Instrumentation II. p. 84503U, \mndoi{10.1117/12.927080}

\bibitem[\protect\citeauthoryear{{Ginsburg} et~al.,}{{Ginsburg}
  et~al.}{2019}]{astroquery}
{Ginsburg} A.,  et~al., 2019, \mndoi [\aj] {10.3847/1538-3881/aafc33}, \href
  {http://adsabs.harvard.edu/abs/2019AJ....157...98G} {157, 98}

\bibitem[\protect\citeauthoryear{{Gisler}, {Feller}  \& {Gandorfer}}{{Gisler}
  et~al.}{2003}]{gisler03}
{Gisler} D.,  {Feller} A.,   {Gandorfer} A.~M.,  2003, in {Fineschi} S.,  ed.,
  \procspie Vol. 4843, Polarimetry in Astronomy. pp 45--54,
  \mndoi{10.1117/12.458835}

\bibitem[\protect\citeauthoryear{{Guthrie}}{{Guthrie}}{1987}]{guthrie87}
{Guthrie} B.~N.~G.,  1987, Quarterly Journal of the Royal Astronomical Society,
  \href {https://ui.adsabs.harvard.edu/\#abs/1987QJRAS..28..289G} {28, 289}

\bibitem[\protect\citeauthoryear{{Horton} et~al.,}{{Horton}
  et~al.}{2012}]{horton12}
{Horton} A.,  et~al., 2012, in \procspie. p. 84463A (\mn@eprint {arXiv}
  {1301.0670}), \mndoi{10.1117/12.924945}

\bibitem[\protect\citeauthoryear{{Hough}, {Lucas}, {Bailey}, {Tamura}, {Hirst},
  {Harrison}  \& {Bartholomew-Biggs}}{{Hough} et~al.}{2006}]{hough06}
{Hough} J.~H.,  {Lucas} P.~W.,  {Bailey} J.~A.,  {Tamura} M.,  {Hirst} E.,
  {Harrison} D.,   {Bartholomew-Biggs} M.,  2006, \mndoi [\pasp]
  {10.1086/507955}, \href {http://adsabs.harvard.edu/abs/2006PASP..118.1302H}
  {118, 1302}

\bibitem[\protect\citeauthoryear{{Hsu} \& {Breger}}{{Hsu} \&
  {Breger}}{1982}]{hsu82}
{Hsu} J.~C.,  {Breger} M.,  1982, \mndoi [\apj] {10.1086/160467}, \href
  {https://ui.adsabs.harvard.edu/\#abs/1982ApJ...262..732H} {262, 732}

\bibitem[\protect\citeauthoryear{Hunter}{Hunter}{2007}]{matplotlib}
Hunter J.~D.,  2007, Computing in science \& engineering, 9, 90

\bibitem[\protect\citeauthoryear{Jones, Oliphant, Peterson  et~al.}{Jones
  et~al.}{2001}]{scipy}
Jones E.,  Oliphant T.,  Peterson P.,   et~al., 2001, {SciPy}: Open source
  scientific tools for {Python}, \url {http://www.scipy.org/}

\bibitem[\protect\citeauthoryear{{Kemp} \& {Barbour}}{{Kemp} \&
  {Barbour}}{1981}]{kemp81}
{Kemp} J.~C.,  {Barbour} M.~S.,  1981, \mndoi [\pasp] {10.1086/130870}, \href
  {http://adsabs.harvard.edu/abs/1981PASP...93..521K} {93, 521}

\bibitem[\protect\citeauthoryear{{Lucas}, {Hough}, {Bailey}, {Tamura}, {Hirst}
  \& {Harrison}}{{Lucas} et~al.}{2009}]{lucas09}
{Lucas} P.~W.,  {Hough} J.~H.,  {Bailey} J.~A.,  {Tamura} M.,  {Hirst} E.,
  {Harrison} D.,  2009, \mndoi [\mnras] {10.1111/j.1365-2966.2008.14182.x},
  \href {http://adsabs.harvard.edu/abs/2009MNRAS.393..229L} {393, 229}

\bibitem[\protect\citeauthoryear{{Marshall} et~al.,}{{Marshall}
  et~al.}{2016}]{marshall16}
{Marshall} J.~P.,  et~al., 2016, \mndoi [\apj] {10.3847/0004-637X/825/2/124},
  \href {http://adsabs.harvard.edu/abs/2016ApJ...825..124M} {825, 124}

\bibitem[\protect\citeauthoryear{{Martin}, {Clayton}  \& {Wolff}}{{Martin}
  et~al.}{1999}]{martin99}
{Martin} P.~G.,  {Clayton} G.~C.,   {Wolff} M.~J.,  1999, \mndoi [\apj]
  {10.1086/306613}, \href
  {https://ui.adsabs.harvard.edu/\#abs/1999ApJ...510..905M} {510, 905}

\bibitem[\protect\citeauthoryear{{McDavid}}{{McDavid}}{2000}]{mcdavid00}
{McDavid} D.,  2000, \mndoi [\aj] {10.1086/301186}, \href
  {https://ui.adsabs.harvard.edu/\#abs/2000AJ....119..352M} {119, 352}

\bibitem[\protect\citeauthoryear{{Morris} et~al.,}{{Morris}
  et~al.}{2018}]{astroplan}
{Morris} B.~M.,  et~al., 2018, \mndoi [\aj] {10.3847/1538-3881/aaa47e}, \href
  {http://adsabs.harvard.edu/abs/2018AJ....155..128M} {155, 128}

\bibitem[\protect\citeauthoryear{{Nakamura}, {Hamana}, {Ishigami}  \&
  {Matsui}}{{Nakamura} et~al.}{2010}]{nakamura10}
{Nakamura} K.,  {Hamana} Y.,  {Ishigami} Y.,   {Matsui} T.,  2010, \mndoi
  [Nuclear Instruments and Methods in Physics Research A]
  {10.1016/j.nima.2010.02.220}, \href
  {http://adsabs.harvard.edu/abs/2010NIMPA.623..276N} {623, 276}

\bibitem[\protect\citeauthoryear{Oliphant}{Oliphant}{2006}]{numpy}
Oliphant T.~E.,  2006, A guide to NumPy.
 Vol. 1, Trelgol Publishing USA

\bibitem[\protect\citeauthoryear{{Patriarchi}, {Morbidelli}, {Perinotto}  \&
  {Barbaro}}{{Patriarchi} et~al.}{2001}]{patriarchi01}
{Patriarchi} P.,  {Morbidelli} L.,  {Perinotto} M.,   {Barbaro} G.,  2001,
  \mndoi [\aap] {10.1051/0004-6361:20010496}, \href
  {https://ui.adsabs.harvard.edu/\#abs/2001A&A...372..644P} {372, 644}

\bibitem[\protect\citeauthoryear{{Piirola}, {Berdyugin}  \&
  {Berdyugina}}{{Piirola} et~al.}{2014}]{piirola14}
{Piirola} V.,  {Berdyugin} A.,   {Berdyugina} S.,  2014, in Ground-based and
  Airborne Instrumentation for Astronomy V. p. 91478I,
  \mndoi{10.1117/12.2055923}

\bibitem[\protect\citeauthoryear{{Serkowski}}{{Serkowski}}{1974}]{serkowski74}
{Serkowski} K.,  1974, in {Gehrels} T.,  ed., IAU Colloq. 23: Planets, Stars,
  and Nebulae: Studied with Photopolarimetry. p.~135

\bibitem[\protect\citeauthoryear{{Serkowski}, {Mathewson}  \&
  {Ford}}{{Serkowski} et~al.}{1975}]{serkowski75}
{Serkowski} K.,  {Mathewson} D.~S.,   {Ford} V.~L.,  1975, \mndoi [\apj]
  {10.1086/153410}, \href {http://adsabs.harvard.edu/abs/1975ApJ...196..261S}
  {196, 261}

\bibitem[\protect\citeauthoryear{{Vucina}, {Boccas}, {Araya}  \&
  {Ahhee}}{{Vucina} et~al.}{2006}]{vucina06}
{Vucina} T.,  {Boccas} M.,  {Araya} C.,   {Ahhee} C.,  2006, in Society of
  Photo-Optical Instrumentation Engineers (SPIE) Conference Series. p. 62730W,
  \mndoi{10.1117/12.670866}

\bibitem[\protect\citeauthoryear{{Wiktorowicz} \& {Matthews}}{{Wiktorowicz} \&
  {Matthews}}{2008}]{wiktorowicz08}
{Wiktorowicz} S.~J.,  {Matthews} K.,  2008, \mndoi [\pasp] {10.1086/595966},
  \href {http://adsabs.harvard.edu/abs/2008PASP..120.1282W} {120, 1282}

\bibitem[\protect\citeauthoryear{{Wiktorowicz} \& {Nofi}}{{Wiktorowicz} \&
  {Nofi}}{2015}]{wiktorowicz15}
{Wiktorowicz} S.~J.,  {Nofi} L.~A.,  2015, \mndoi [\apjl]
  {10.1088/2041-8205/800/1/L1}, \href
  {http://adsabs.harvard.edu/abs/2015ApJ...800L...1W} {800, L1}

\bibitem[\protect\citeauthoryear{{Wilking}, {Lebofsky}, {Martin}, {Rieke}  \&
  {Kemp}}{{Wilking} et~al.}{1980}]{wilking80}
{Wilking} B.~A.,  {Lebofsky} M.~J.,  {Martin} P.~G.,  {Rieke} G.~H.,   {Kemp}
  J.~C.,  1980, \mndoi [\apj] {10.1086/157694}, \href
  {https://ui.adsabs.harvard.edu/\#abs/1980ApJ...235..905W} {235, 905}

\bibitem[\protect\citeauthoryear{{Wilking}, {Lebofsky}  \& {Rieke}}{{Wilking}
  et~al.}{1982}]{wilking82}
{Wilking} B.~A.,  {Lebofsky} M.~J.,   {Rieke} G.~H.,  1982, \mndoi [\aj]
  {10.1086/113147}, \href {http://adsabs.harvard.edu/abs/1982AJ.....87..695W}
  {87, 695}

\makeatother
\end{thebibliography}

\appendix

\section{Modulation efficiency at high and low polarization}
\label{sec:apa}
As mentioned in section \ref{sec:bandpass} the modulation efficiency measured for the instrument is different when illuminated with 100\% polarized light in the laboratory and when observing astronomical sources with low polarizations. The difference arises because of the way the polarization alters the average intensity of the two beams when the modulator departs from the ideal half-wave retardance.

The instrument is essentially a retarder (the FLC modulator) followed by a polarizer (the Wollaston prism). The Mueller matrix for a retarder is \citep{bailey15}:

\begin{equation}
\begin{bmatrix} 1 & 0 & 0 & 0 \\ 
0 & C^2+S^2 \cos{\delta} & SC(1-\cos{\delta}) & -S \sin{\delta} \\
0 & SC(1-\cos{\delta}) & S^2 + C^2\cos{\delta} & C \sin{\delta} \\
0 & S \sin{\delta} & -C\sin{\delta} & \cos{\delta} 
\end{bmatrix}
\end{equation}
where $C = \cos{2 \phi}$, $S = \sin{2 \phi}$, $\delta$ is the retardance, and $\phi$ the angle 
to the fast axis of the retarder. For a half wave plate the retardance is
$\pi$ radians.

The Mueller matrix for the polarizer is:

\begin{equation}
1/2 \begin{bmatrix} 1 & c & s & 0 \\
c & c^2 & sc & 0 \\
s & cs & s^2 & 0 \\
0 & 0 & 0 & 0 
\end{bmatrix}
\end{equation}
where $c = \cos{2 \theta}$, $s = \sin{2 \theta}$, 
$\theta$ is the angle of the polarizer axis to that defined for the incoming beam. 

Consider the effect of these optical elements on a polarized input beam with Stokes vector \{1, q, 0, 0\}. Multiplying by the top two rows of the retarder matrix gives:

\begin{equation}
    I = 1
\end{equation}
\begin{equation}
    Q = (C^2 + S^2 \cos{\delta}) q
\label{eqn_a1}    
\end{equation}
for the I and Q Stokes parameters after the retarder. If the two orientations of the fast axis of the modulator for the two modulation states are $\phi = 0$ degrees ($C = 1, S = 0$) and $\phi = 45$ degrees ($C = 0, S = 1$) then equation \ref{eqn_a1} becomes:

\begin{equation}
    Q_1 = q 
\end{equation}
\begin{equation}
    Q_2 = q \cos{\delta}
\end{equation}

for the two modulation states and we can then multiply by the top row of the polarizer matrix to get the output intensity in each state (assuming $\theta = 0$ and hence $c = 1$):

\begin{equation}
    I_1 = 0.5 (1 + q)
\end{equation}
\begin{equation}
    I_2 = 0.5 (1 + q \cos{\delta})
\end{equation}

The intensities in the second Wollaston beam for which we can assume $\theta = 90$ degrees and hence $c = -1$ are the same equations with a minus sign replacing the plus sign. The modulation that we measure is given by:

\begin{equation}
    M_A = \frac{I_1 - I_2}{I_1 + I_2} = \frac{q (1-\cos{\delta})}{2+q+q \cos\delta}
    \label{eqn_ma1}
\end{equation}

and the equivalent for the second Wollaston beam is:

\begin{equation}
    M_B = \frac{I_1 - I_2}{I_1 + I_2} = \frac{-q (1-\cos{\delta})}{2-q-q \cos\delta}
    \label{eqn_mb1}
\end{equation}

The minus sign here meaning that the modulation in the two Wollaston beams are of opposite signs. The modulation efficiency is the modulation divided by the input polarization (i.e.  $M_A/q$ and $M_B/q$).

We can now consider several special cases of the general formulae in equations \ref{eqn_ma1} and \ref{eqn_mb1}. If the modulator is a half-wave retarder then $\cos{\delta} = -1$ and therefore
\begin{equation}
    M_A/q = 1  
\end{equation}
\begin{equation}
    M_B/q = -1
\end{equation}
This is the ideal case giving 100\% modulation efficiency.

If $q$ is very much less than one (i.e. low polarization) then we can ignore the $q$ and $q \cos{\delta}$ terms in the denominator and we get:

\begin{equation}
    M_A/q = \frac { 1 - \cos{\delta}}{2}  
\end{equation}
\begin{equation}
    M_B/q = -\frac { 1 - \cos{\delta}}{2} 
\end{equation}

which is the form used in equation \ref{eq:e_lp}.

If $q = 1$ (i.e. 100\% input polarization as in our laboratory calibration) then we get 

\begin{equation}
    M_A/q = \frac { 1 - \cos{\delta}}{3 + \cos{\delta}}  
\end{equation}
\begin{equation}
    M_B/q = -\frac { 1 - \cos{\delta}}{1 - \cos{\delta}} = -1
\end{equation}

In this case the modulation efficiencies for the two beams have different magnitudes as well as opposite signs. Averaging the magnitude of these two gives the expression used in equation \ref{eq:e_hp}.

In other cases where the polarization is large, equations \ref{eqn_ma1} and \ref{eqn_mb1} must be used.

\end{document}